\documentclass[aps,prx,superscriptaddress,twocolumn,showpacs,longbibliography,nofootinbib]{revtex4-1}

\usepackage{amsfonts}
\usepackage{hyperref}
\usepackage[usenames,dvipsnames]{xcolor}
\usepackage{amsmath,amsthm,amssymb,dsfont}
\usepackage{enumerate}
\usepackage[english]{babel}
\usepackage{graphicx}
\usepackage[caption=false]{subfig}
\usepackage{bbm}
\usepackage{epsfig}
\usepackage{color}
\usepackage{epstopdf}

\usepackage{epsfig}

\usepackage{braket}

\usepackage[normalem]{ulem}

\newcommand{\ketbra}[2]{| #1 \rangle \langle #2 |}

\newcommand{\id}{\mathbb{I}}

\newcommand{\bea}{\begin{eqnarray}}
\newcommand{\eea}{\end{eqnarray}}

\DeclareMathOperator{\tr}{Tr}

\begin{document}
	\widetext
	\title{Experimentally reducing the quantum measurement back-action in work distributions by a collective measurement}
	\date{\today}
	
	\author{Kang-Da Wu}
	\affiliation{CAS Key Laboratory of Quantum Information, University of Science and Technology of China, Hefei, 230026, People's Republic of China}
	\affiliation{CAS Center For Excellence in Quantum Information and Quantum Physics, University of Science and Technology of China, Hefei, 230026, People's Republic of China}
	
	\author{Yuan Yuan}
	\affiliation{CAS Key Laboratory of Quantum Information, University of Science and Technology of China, Hefei, 230026, People's Republic of China}
	\affiliation{CAS Center For Excellence in Quantum Information and Quantum Physics, University of Science and Technology of China, Hefei, 230026, People's Republic of China}

	\author{Guo-Yong Xiang}
	\email{gyxiang@ustc.edu.cn}
	\affiliation{CAS Key Laboratory of Quantum Information, University of Science and Technology of China, Hefei, 230026, People's Republic of China}
	\affiliation{CAS Center For Excellence in Quantum Information and Quantum Physics, University of Science and Technology of China, Hefei, 230026, People's Republic of China}

	\author{Chuan-Feng Li}
	\affiliation{CAS Key Laboratory of Quantum Information, University of Science and Technology of China, Hefei, 230026, People's Republic of China}
	\affiliation{CAS Center For Excellence in Quantum Information and Quantum Physics, University of Science and Technology of China, Hefei, 230026, People's Republic of China}

	\author{Guang-Can Guo}
	\affiliation{CAS Key Laboratory of Quantum Information, University of Science and Technology of China, Hefei, 230026, People's Republic of China}
	\affiliation{CAS Center For Excellence in Quantum Information and Quantum Physics, University of Science and Technology of China, Hefei, 230026, People's Republic of China}
	
	\author{Mart\'{\i} Perarnau-Llobet}
	\email{marti.perarnau@mpq.mpg.de}
	\affiliation{Max-Planck-Institut f\"{u}r Quantenoptik, Hans-Kopfermann-Str. 1, D-85748 Garching, Germany}
	
	\begin{abstract}
		In quantum thermodynamics, the standard approach to estimate work fluctuations in unitary processes is based on two projective measurements, one performed at the beginning of the process and one at the end. The first measurement destroys any initial coherence in the energy basis, thus preventing later interference effects. In order to decrease this back-action, a scheme  based on collective measurements has been proposed in~[PRL 118, 070601 (2017)]. Here, we report its experimental implementation in an optical system. The experiment consists of a deterministic collective measurement on identically prepared two qubits, encoded in the polarisation and path degree of a single photon.  The standard two projective measurement approach is also experimentally realized for comparison. Our results show the potential of collective schemes to decrease the back-action of projective measurements, and capture subtle effects arising from quantum coherence.
	\end{abstract}
	\maketitle
	
	\section{Introduction}
	
	Quantum coherence lies at the heart of quantum physics. Yet, its presence is subtle to observe, as projective measurements inevitably destroy it. 
	In the context of quantum thermodynamics, this tension becomes apparent in work fluctuations: Whereas projective energy measurements are commonly employed to measure them~\cite{TPM1RMP,TPM2PRE}, they also lead to work  distributions that are independent of the initial coherence in the energy basis.
	This limitation has motivated alternative proposals for defining and measuring
	work in purely coherent evolutions~\cite{Allahverdyan2014nonequilibrium,
		Solinas2015full,dechiara2015measuring,Talkner2016aspects,Solinas2016g,de2018ancilla,Deffner2016quantum,Miller2017time,Sampaio2017impossilbe,xu2017effects,Solinas2013work,Solinas2016probing,
		Solinas2017measurement,Hofer2017quasiprobability,Lostaglio2017quantum,NoGoTheorem,baumer2018fluctuating},  which include Gaussian~\cite{dechiara2015measuring,Talkner2016aspects,Solinas2016g,de2018ancilla}, weak   \cite{Solinas2017measurement,Hofer2017quasiprobability,Lostaglio2017quantum}, and collective measurements~\cite{NoGoTheorem}.
	These different theoretical proposals aim at reducing the back-action induced by projective measurements, thus allowing for preserving some coherent interference effects. This quest is particularly relevant as, when the system is left unobserved, quantum coherence can play an important role in several thermodynamic tasks; e.g. in work extraction~\cite{korzekwa2016extraction,Patrick2018} and heat engines~\cite{Rossnagel2014nanoscale,Correa2014quantum,Mitchison2015Coherence,Klatzow2017Experimental}.
	Indeed, quantum coherence can be seen as a source of free energy, which is destroyed by projective energy measurements \cite{lostaglio2015description,Janzing2006}.

	In this article, we report the experimental investigation of reducing quantum measurement back-action in work distribution by using collective measurements~(CM) on two identically prepared qubit states. We implement the proposal of~\cite{NoGoTheorem} in an all-optical setup, which can be used to efficiently simulate quantum coherent processes. The standard two projective energy measurement (TPM) scheme~\cite{TPM1RMP,TPM2PRE} to measure work is also experimentally simulated for comparison. The experimental results show the capability of CM to capture coherent effects and reduce the measurement back-action, which is quantified as the fidelity between the probability distributions of the final measured and unmeasured states.  
	
    Moreover, the potential application of these results goes  beyond quantum thermodynamics, as deterministic collective measurements play a key role in quantum information, being relevant for numerous tasks such as quantum metrology~\cite{Giov11advances,bohnet2014reduced}, tomography~\cite{Massar1995,hou2018deterministic}, and state manipulation~\cite{Cox2016}. \\

	
	\begin{figure*}[htp]
		\label{fig:exp}
		\includegraphics[scale=0.18]{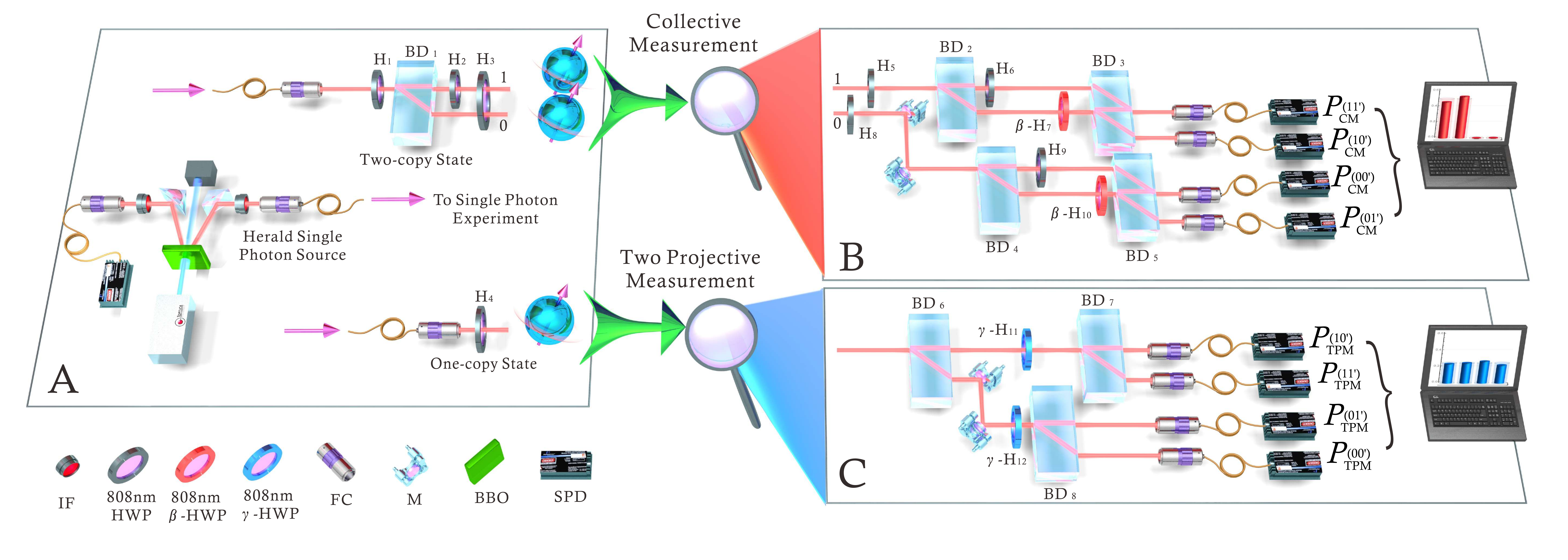}
		\caption{\label{fig:exp} \textbf{Experimental Set-up for both CM scheme and TPM Scheme.}
			The set up is divided in three modules: State Preparation (A), Collective Measurements (B) and Two Projective Measurement (C). Module A can generate an arbitrary one-copy polarization encoded state $\ket{\Phi}$ or a two-copy polarization-path encoded state $\ket{\Phi}^{\otimes2}$ of a single photon. Module B implements the CM on $\ket{\Phi}^{\otimes2}$. The rotation angle of two $\beta$-HWPs are adjustable for different unitary processes $U(\theta)$ with $\cos^22\beta=2\sin^2\theta$. The rotation angles of the other four HWPs are fixed as: H$_5$-$22.5^{\circ}$; H$_6$-$45^{\circ}$; H$_8$-$67.5^{\circ}$; H$_9$-$45^{\circ}$. Module C implements the TPM schemes on $\ket{\Phi}$, the rotation angle of the two $\gamma$-HWPs is adjustable and can implement different $U(\theta)$ with $\theta=2\gamma$. Key to components: SPD, single photon detector; HWP, half-wave plate; FC, fiber coupler; BD, beam displacer; M, mirror; BBO, $\beta$-barium borate; IF, interference filter.}
	\end{figure*}

	\section{Results}
	\subsection{Theoretical Framework}
	
	The scenario considered here consists of a quantum state $\rho$ and a Hamiltonian $H$. The system is taken to be thermally isolated, and it can only be modified by externally driving $H$. We consider processes in which $H$ is transformed up to $H'$, and as a consequence the state evolves under a unitary evolution $U$, $\rho\rightarrow U\rho U^{\dagger}$.
	The average energy for this process is given by,
	\begin{align}
	\langle W \rangle =\tr(H \rho)-\tr(H' U\rho U^{\dagger}),
	\label{unmeasuredwork}
	\end{align}
	the energy difference can be identified with unmeasured average work.
	However, when one attempts to measure it, the average measured work usually differs from Eq.~\eqref{unmeasuredwork} due to measurement back-action~\cite{TPM1RMP,Allahverdyan2014nonequilibrium,
		Talkner2016aspects,NoGoTheorem,Kammerlander2016quantum}. 
	
	In the standard approach to measure work in quantum systems~\cite{TPM1RMP,TPM2PRE}, one implements two energy measurements, of $H$ and $H'$, before and after the evolution $U$. More precisely, expanding the Hamiltonians in the bra-ket representation, as $H=\sum_iE_i\ketbra{i}{i}$ and $H'=\sum_{j'} E'_{j'}\ketbra{j'}{j'}$, the TPM consists of 
		\begin{itemize}
			\item Projective measurement of $H$ on $\rho$ yielding outcome  $E_i$ with probability $\rho_{ii} =\bra{i}\rho \ket{i}$.
			\item A unitary evolution $U$ of the postmeasured state, $\ket{i} \rightarrow U \ket{i}$.
			\item A projective measurement of $H'$ on the evolved state, yielding $E_j'$ with probability $p_{i,j'}=|\bra{j'}U \ket{i}|^2$. 
		\end{itemize}  
		The TPM work statistics are then given by the random variable $w^{(ij')}=E_i-E'_{j'}$
		with a corresponding probability $P_{\rm TPM}^{(ij')}=\rho_{ii} p_{i,j'}$ assigned to the transition $\ket{i}\rightarrow\ket{j'}$.
		The average measured work,  $\langle W_{\rm TPM}\rangle \equiv \sum_{ij} P_{\rm TPM}^{(ij')} w^{(ij')}$, can be written as 
	\begin{align}
	\langle W_{\rm TPM}\rangle =  \tr(H \mathcal{D}_H[\rho])- \tr(H' U \mathcal{D}_H [\rho] U^{\dagger}),
	\label{avTPM}
	\end{align}
	where $\mathcal{D}_H[\rho]$ is the dephasing operator, removing all the coherence of $\rho$, which yields a classical mixture of energy states of $H$. Hence, $\langle W_{\rm TPM}\rangle$ differs from the unmeasured average work in Eq.~\eqref{unmeasuredwork} when $\rho$ is coherent (and $[H,U^{\dagger}H'U]\neq 0$). Furthermore, the extractable work
		from $\mathcal{D}_H[\rho]$ is lower than that of $\rho$, as the latter is in general more pure. This can be seen by noting that the non-equilibrium free energy, which characterises the extractable work from a state,  decomposes into a contribution arising from $\mathcal{D}_H[\rho]$ and one from the coherent part of   $\rho$ \cite{lostaglio2015description,Janzing2006} (see also 
		Appendix A of \cite{baumer2018fluctuating}).
	
	In order to reduce the back-action  of the TPM scheme, a collective measurement (CM) has been proposed~\cite{NoGoTheorem}. To describe such measurements, let us now introduce the formalism of generalised measurements, which extends the standard quantum projective measurements.  A generalised measurement is defined by a positive-operator valued measure (POVM) \cite{nielsen2002quantum}, which is a set of non-negative Hermitian operators $\{M^{(i)}\}$ satisfying the completeness condition $\sum_{\{ i \}} M^{(wi)} = \mathbb{I}$.  Each operator $M^{(i)}$ is associated to a   measurement outcome  $w^{(i)}$ of the experiment. Then, given a quantum state $\rho$, the probability to obtain the $w^{(i)}$ is given by the generalized Born rule:
		\begin{equation}
		P^{(i)}=\tr\left(\rho M^{(i)}\right).
		\label{Pi}
		\end{equation}
		Note that the completeness condition ensures that the sum of probability obtained from each outcome $i$ is equal to 1. 
		Collective measurements (CM) can then be naturally introduced by taking $\rho$ to be a collection of $n$ independent systems,  $\rho=\rho_1 \otimes \rho_2 \otimes ... \otimes \rho_n$, so that
		\begin{equation}
		P^{(i)}=\tr\left(\rho_1 \otimes \rho_2 \otimes ... \otimes \rho_n M^{(i)}\right).
		\label{Pi}
		\end{equation}
		That is, the measurement  acts globally on the $n$ systems.
	In this work, we consider systems made up of two qubits, so that the collective measurements act globally on a Hilbert space of four dimensions.
	
	At this point, it is useful to express the TPM scheme as a POVM, with elements $M^{(ij')}_{\rm TPM}=\big|\hspace{-1mm}\bra{j'}U \ket{i}\hspace{-1mm}\big|^2 \ketbra{i}{i}$ and probability assigned $P_{\rm TPM}^{(ij')}= \tr(M^{(ij')}_{\rm TPM}  \rho)$, where $\ketbra{i}{i}$ denotes a projection on energy basis $\ket{i}$.	On the other hand, the CM scheme is defined by a POVM with elements $M^{(ij')}_{\rm CM}$ that act  on two copies of the state, $\rho^{\otimes 2}$, with associated probability $P_{\rm CM}^{(ij')}= \tr(M^{(ij')}_{\rm CM} \rho^{\otimes 2})$. The POVM elements read, 
	\begin{align}
	M^{(ij')}_{\rm CM}= M^{(ij')}_{\rm TPM} \otimes \id + \lambda \ketbra{i}{i} \otimes T_{j'}^{\rm off-diag},
	\label{M^{(ij)}}
	\end{align}
	where $T^{\rm off-diag}_{j'}$ is the off-diagonal part of $T_{j'}=U^{\dagger} \ketbra{j'}{j'} U$ in the $\{ \ket{i} \}$ basis. This  measurement satisfies two basic properties
	\begin{enumerate}
		\item When acting upon states with zero coherence, $\rho=\mathcal{D}(\rho)$, the CM scheme reproduces exactly the same statistics of the standard TPM scheme. This follows by noting that $\tr(T_{j'}^{\rm off-diag} \mathcal{D}(\rho))=0$ and $\tr(\mathcal{D}(\rho))=1$ in Eq.~\eqref{M^{(ij)}}.
		\item When acting upon general $\rho$, the second term of Eq.~\eqref{M^{(ij)}} brings information about the purely coherent part of the evolution. This can be seen by computing the average measured work, $\langle W_{\rm CM}\rangle=\sum_{i,j'}w^{(ij')}P_{\rm CM}^{(ij')}$, leading to
		\begin{align}\label{averWCM}
		\langle W_{\rm CM}\rangle=(1-\lambda) \langle W_{\rm TPM}\rangle + \lambda \langle W \rangle. 
		\end{align}
	\end{enumerate}
	Hence, the parameter $\lambda \in [0,1]$ quantifies the degree of  measurement back-action.
	In general, $\lambda$ is given by an optimisation procedure, which is described in the Methods, and it can be controlled in our experiment.
	We also note that other proposals of work measurements in states with quantum coherence, in particular weak or gaussian measurements, can interpolate between Properties 1. and 2 described above. In the limit of strong (weak) measurements, Property 1. (2.) is satisfied, whereas for intermediate couplings with the apparatus a tradeoff appears  (see \cite{dechiara2015measuring,Talkner2016aspects,Solinas2016g,de2018ancilla} for discussions).
	

	
With the probabilities $P^{(ij')}$, which can be obtained either by the TPM or the CM schemes, the full work distribution is constructed as 
		\begin{align}
		P(w) = \sum_{ij} P^{(ij')} \delta(w-w^{(ij')})
		\label{P(w)}
		\end{align}
		where $\delta$ is a Dirac delta function, which accounts for possible degeneracies in  $w^{(ij)}$. \\

    \subsection{Experimental Protocol}
	
	We consider the experimental realisation of the CM in Eq.~\eqref{M^{(ij)}} on a two-qubit syste in a quantum optics set-up. The core idea is to encode the first (second) copy into the path (polarisation) degree of freedom of a single photon, as illustrated in Fig. \ref{fig:exp}. Single photons have degenerate Hamiltonians for both polarisation and path degree, i.e.,  $w^{(ij)}=0$ for all $i,j$, leading to a priori trivial work distributions $P(w)$ in Eq.~\eqref{P(w)}. Yet,  $P(w)$ is a coarse-grained version of the transition probabilities $P^{(ij')}$, and the latter contains all information about the quantum stochastic process. Therefore, we focus on $P^{(ij')}$, and attempt to capture the subtle effect of quantum coherence in the process by working on the experimentally highly non-trivial two-copy space.

	We consider unitary process of the form $U(\theta)=\cos\theta\sigma_z+\sin\theta\sigma_x$, where $\sigma_x$ and $\sigma_z$ are Pauli operators and the parameter $\theta$ is tunable. For such  $U(\theta)$'s,  we have
	$\lambda=\tan \theta$ ($\theta \in [0, \pi/4]$), leading to
	\begin{equation}
	\begin{aligned}
	&M^{(00')}_{\textmd{CM}}=\ketbra{0}{0}\otimes(\cos^2\theta\mathbb{I}+\sin^2\theta\sigma_{x}),\\
	&M^{(01')}_{\textmd{CM}}=2\sin^2\theta\ketbra{0}{0}\otimes\ketbra{-}{-}, \\
	&M^{(10')}_{\textmd{CM}}=2\sin^2\theta\ketbra{1}{1}\otimes\ketbra{+}{+}, \\
	&M^{(11')}_{\textmd{CM}}=\ketbra{1}{1}\otimes(\cos^2\theta\mathbb{I}-\sin^2\theta\sigma_{x}),
	\end{aligned} 
	\end{equation}
	with $\ket{\pm}=(\ket{0}\pm\ket{1})/\sqrt{2}$. 
	These measurement operators $M^{(ij')}_{\textmd{CM}}$,  associated to the transitions $\ket{i}\rightarrow\ket{j'}$, are the ones implemented in the experiment (together with the TPM scheme).  \\
	
	\subsection{Experimental Setup}
	
	The whole experimental set up is illustrated in Fig.~\ref{fig:exp} and can be divided into three modules: State Preparation module (A), Collective Measurement module (B) and TPM module (C).
	
	In module A, a single-photon state is generated through a type-II beamlike
	phase-matching $\beta$-barium borate (BBO) crystal pumped by a 80 mW continuous wave laser (with a central wave length of 404 nm) via spontaneous parametric down-conversion (SPDC)~\cite{takeuchi2001beamlike}. The initial state can be written as $\ket{0}^{\otimes 2}$, with the first (second) state encoding the path (polarisation) of the photon. Then, the combined action of BD$_1$ and H$_{1,2,3}$ transforms the initial state into a  two-copy state $\ket{\Phi}^{\otimes2}$, with 
	\begin{equation}\label{pure1}
	\ket{\Phi}=\sqrt{p_0}\ket{0}+\sqrt{p_1}\ket{1}
	\end{equation}
	where $p_{0}(p_1)$ is tunable in our experiments, denoting the population of photons initialised in state $\ket{0}(\ket{1})$, and $p_0+p_1=1$. Details of this transformation are provided in the Supplementary Material. Module A also allows for the generation of  a one-copy qubit state in Eq.~(\ref{pure1}), which is fed into the TPM measurement.
	
	The CM scheme is deterministically realized in Module B of Fig.~\ref{fig:exp}. When $\ket{\Phi}^{\otimes2}$ enters the CM module, the projector $\ketbra{i}{i}$ ($i=0,1$) in Eq.~\eqref{M^{(ij)}} on the first copy (path-encoded) is implemented. The information obtained is then fed into a two-element-POVM on the second copy (polarisation-encoded). If the outcome of the path measurement reads $0$, the POVM elements on the second copy are $\cos^2\theta\mathbb{I}+\sin^2\theta\sigma_{x}$ and $2\sin^2\theta\ketbra{-}{-}$ with outcomes $00'$ and $01'$; this is done by H$_8$, H$_9$, $\beta$-H$_{10}$, BD$_4$ and BD$_5$.
	Note that $\beta$-H$_{10}$ implements the unitary transformation $U(\theta)$ through a tunable angle $\beta$ satisfying $\cos^22\beta=2\sin^2\theta$. Similarly, if the outcome reads $1$, the POVM elements  $2\sin^2\theta\ketbra{+}{+}$ and $\cos^2\theta\mathbb{I}-\sin^2\theta\sigma_{x}$ are realised through H$_5$, H$_6$, $\beta$-H$_7$, BD$_2$, and BD$_3$ (see Fig.~\ref{fig:exp}). As in the previous case, $\beta$-H$_7$ implements the unitary $U(\theta)$, with arbitrary $\theta$, by setting $\theta$ to $\cos^22\beta=2\sin^2\theta$. See Methods for more details on module B.
	
	A comparative experiment is performed in Module C for simulating TPM scheme. After the preparation of the one-copy state, the polarisation-encoded photon directly enters the TPM measurement, which is conducted by a first polarisation measurement, followed by $\gamma$-H$_{11}$ and $\gamma$-H$_{12}$ implementing the unitary $U(\theta)$ ($\theta=2\gamma$), and finally sequential projections on the polarisation. The parameter $\gamma$ is tunable and set to $\theta=2\gamma$ in order to implement $U(\theta)$.
	Summarising, the four $M_{\text{TPM}}^{(ij)}$ POVM elements can be experimentally realised in this set up, which can simulate coherent processes $U(\theta)$ with arbitrary $\theta$. \\

	\begin{figure}[htp]
		\label{fig:resultSC}
		\includegraphics[scale=0.11]{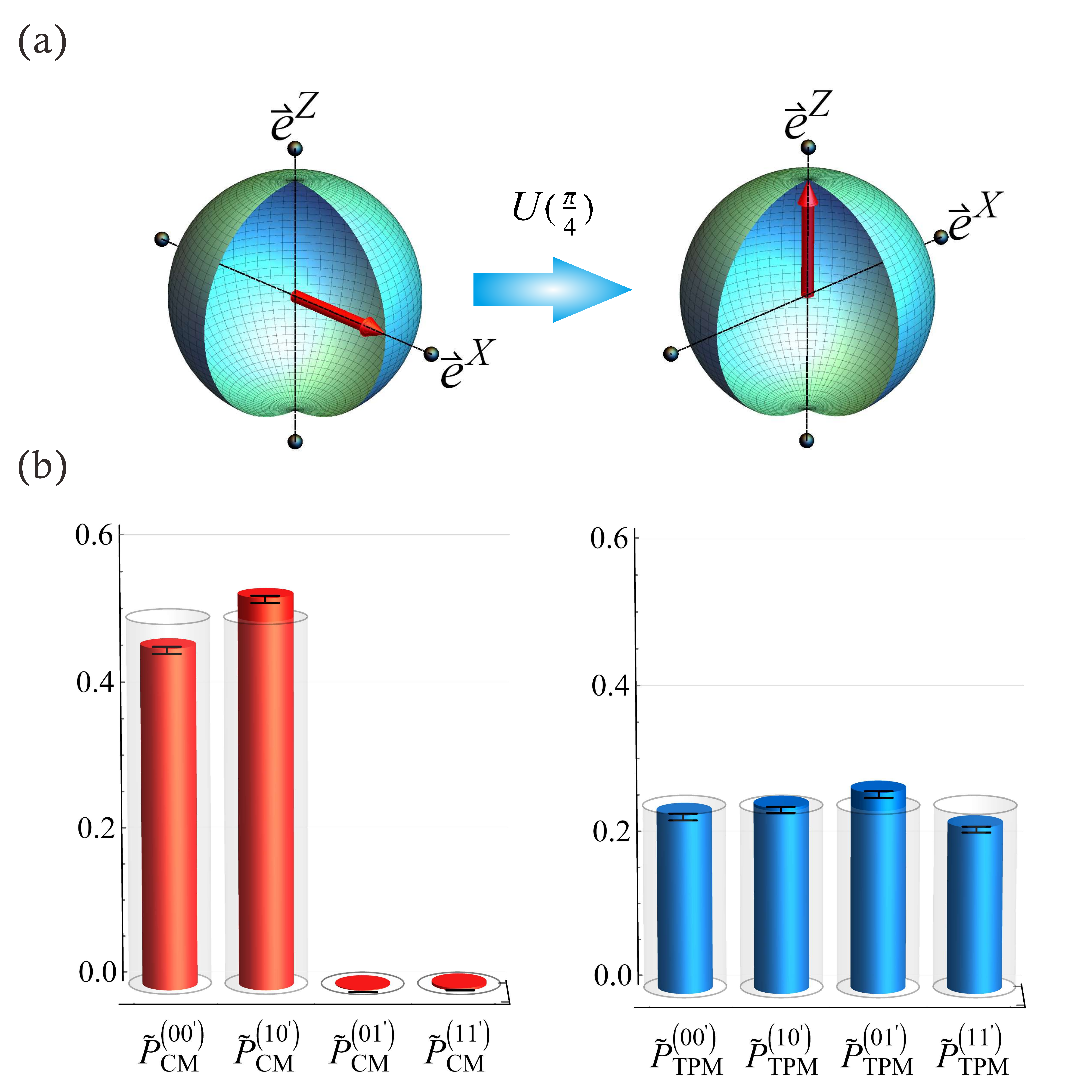}
		\caption{\label{fig:resultSC} \textbf{Transition probabilities for the initial state $\ket{+}$ and  the unitary $U(\pi/4)$ from experimental data.}
			Experimental results for the transition probabilities of the CM and the TPM measurement correspond to the red and blue cylinder, respectively. In (a), the factual transition $U(\pi/4)$ takes an initial maximally coherent state $\ket{+}$ to an incoherent pure state $\ket{0}$, the quantum states are shown by Bloch representation. In (b), the transition probabilities for the CM are $\tilde{P}^{(00')}_{\textmd{CM}}=0.464, \tilde{P}^{(10')}_{\textmd{CM}}=0.532, \tilde{P}^{(10')}_{\textmd{CM}}=0.001, \tilde{P}^{(11')}_{\textmd{CM}}=0.003$. And the results of  the TPM  are  $\tilde{P}^{(00')}_{\textmd{TPM}}=0.244, \tilde{P}^{(10')}_{\textmd{TPM}}=0.254, \tilde{P}^{(10')}_{\textmd{TPM}}=0.275, \tilde{P}^{(11')}_{\textmd{TPM}}=0.227$ respectively. The theoretical fitting values are shown by black edged transparent cylinders.
		}
	\end{figure}
	\begin{figure}[htp]
		\label{fig:resultx}
		\includegraphics[scale=0.11]{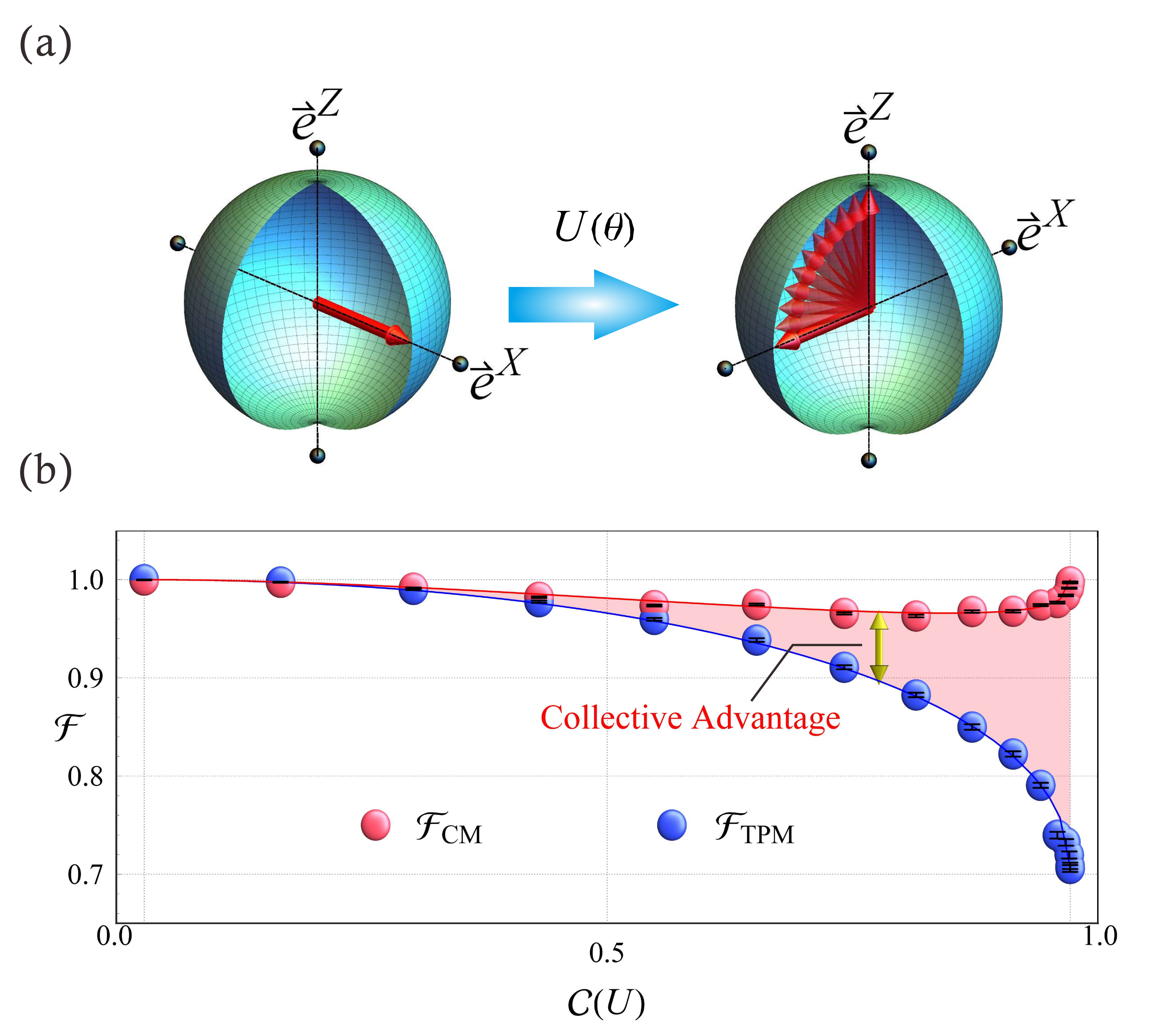}
		\caption{\label{fig:exx} \textbf{Measurement back-action (obtained from experimental data) for various coherent processes.} Experimental results for the measurement back-action, quantified by the fidelity between measured and unmeasured final energy distributions, of the TPM scheme (in blue) and the CM scheme (in red). The results are obtained by fixing the initial state to a maximally coherent state $\ket{+}$ and for various  unitary processes $U(\theta)$  with $\theta$ between $0^\circ$ and $45^\circ$, mapping a fixed input to a class of pure states, as shown in (a). The fidelity in (b) is plotted against the cohering power of $U(\theta)$.
		}
	\end{figure}
	
	\begin{figure}[htp]
		\label{fig:resulty}
		\includegraphics[scale=0.11]{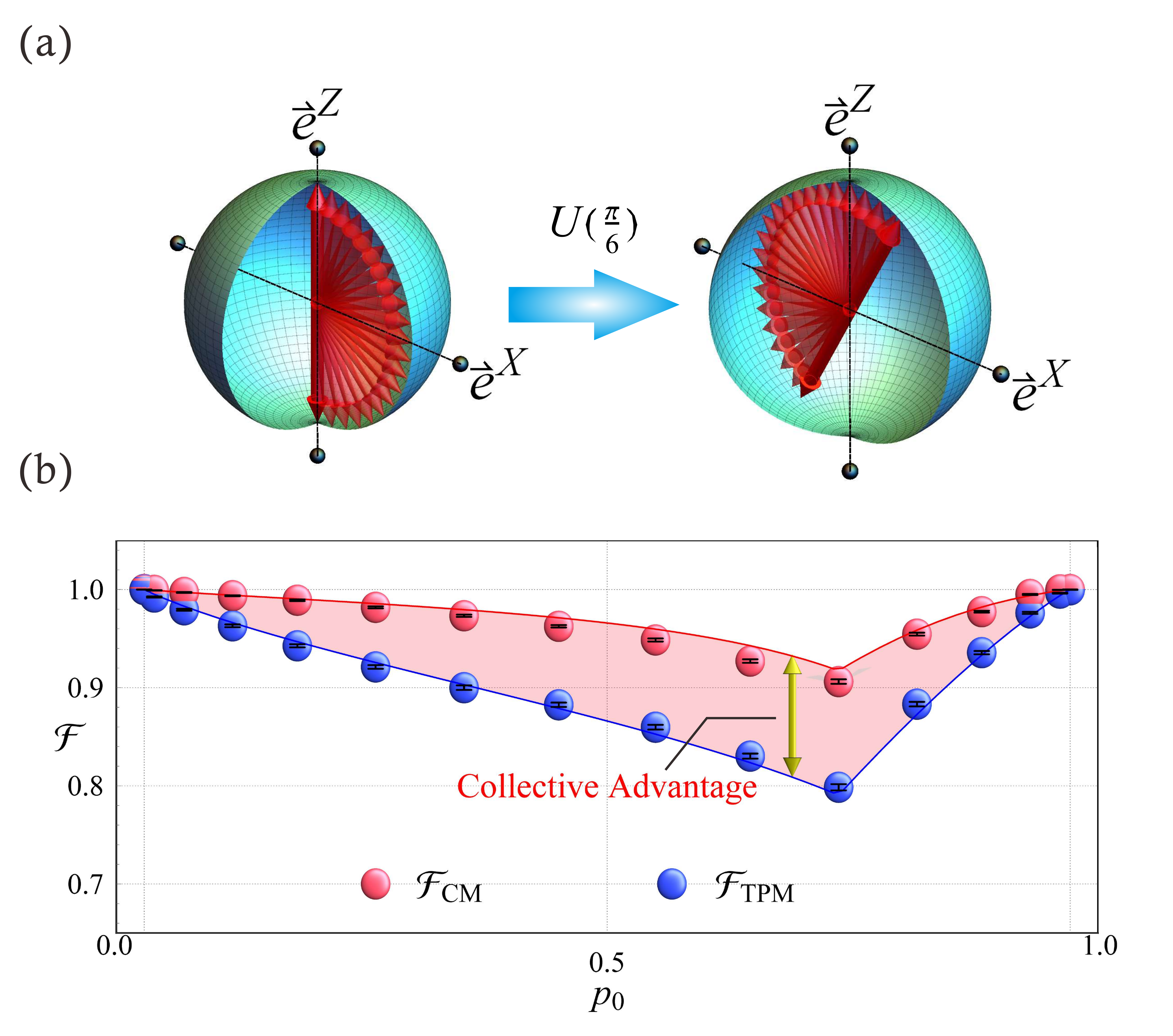}
		\caption{\label{fig:exy} \textbf{Measurement back-action (obtained from experimental data) for different initial states.}
			Experimental results for the fidelity between the measured and unmeasured final energy distributions, for both the TPM scheme (in blue) and the CM scheme (in red). The results are obtained for input states of the form $\ket{\Phi}=\sqrt{p_0}\ket{0}+\sqrt{p_1}\ket{1}$, for various values of $p_0$ between $0$ and $1$, and $p_0+p_1=1$. The unitary is fixed to $U(\pi/6)$ with a cohering power of $\sqrt{3}/2$, mapping a class of pure states to another class of pure states, as shown in (a).  The experimental results in (b) agree well with theoretical predictions.
		}
	\end{figure}

    \subsection{Experimental Results}
	
	We conduct both two schemes for different initial states and unitary processes, with the aim of characterising the measurement back-action.  For characterising coherent states and coherent evolutions, we use l-1 norm coherence $C_{l1}(\rho)$~\cite{QuantifyingCoherence} and cohering power of a unitary $\mathcal{C}(U)$~\cite{coheringpowerPLA}. The l-1 norm coherence measures the degree of interference between different energy basis, and the cohering power quantifies the maximal coherence that can be generated from incoherent states (for more details see the Supplementary Material). 
	
	The experiments are divided into two parts. In the first part, both measurement schemes are implemented on a pure maximally coherent input state $\ket{+}$ undergoing different unitary processes $U(\theta)$. In the second part, we test the above two measurements on various input $\ket{\Phi}$ while fixing the $U(\theta)$.
	
	In order to make a quantitative analysis on the back-action, we compare the probability distributions of ending in state $\ket{j'}$, with $j' =\{0,1\}$, for the unmeasured and measured states -- either by TPM or CM. The strength of the measurement back-action is quantified by the fidelity $\mathcal{F}$ between both distributions, so that for $\mathcal{F}=1$ there is no back-action. The probability distribution of the unmeasured final state can be computed as $P^{(j')}_{\textmd{Id}}=|\bra{j'}U(\theta)\ket{\Phi}|^2$ with $j'=0,1$; whereas the measured final distribution is obtained as $\tilde{P}^{(j')}_{\textmd{CM}}=\sum_{i}\tilde{P}^{(ij')}_{\textmd{CM}}$ and $\tilde{P}^{(j')}_{\textmd{TPM}}=\sum_{i}\tilde{P}^{(ij')}_{\textmd{TPM}}$ for the CM and TPM schemes, respectively, and where the superscript in $\tilde{P}$ indicates that it is obtained from experimental data.
	
	To illustrate our results, we first consider the evolution of $\ket{+}\equiv (\ket{0}+\ket{1})/\sqrt{2}$ towards $\ket{0}$ through $U(\pi/4)$. The measured probabilities $\tilde{P}^{(ij)}$ are shown in Fig.~\ref{fig:resultSC}, plotted
	as red and blue cylinders for the CM and the TPM scheme, respectively. The theoretical values for both schemes are shown with a black edged transparent cylinder. We observe strong differences between the TPM and the CM distributions; being the latter results the ones naively expected from the unmeasured evolution $\ket{+}\rightarrow \ket{0}$. Indeed, the probabilities for ending in $0'$ and $1'$ are given by $\tilde{P}^{(0')}_{\textmd{CM}}=0.996$ and $\tilde{P}^{(1')}_{\textmd{CM}}=0.004$ for the CM,  by $\tilde{P}^{(0')}_{\textmd{TPM}}=0.498$ and $\tilde{P}^{(1')}_{\textmd{TPM}}=0.502$ for the TPM; while the unmeasured evolution yields $P^{(0')}_{\textmd{Id}}=1.0$ and $P^{(1')}_{\textmd{Id}}=0.0$. The fidelity, which measures the back-action, for the above two schemes reads $F_{\textmd{CM}}=0.998$ and $F_{\textmd{TPM}}=0.706$, respectively.

	Experimental results for different coherent processes are shown in Fig.~\ref{fig:exx}. The cohering power is tuned by the rotation angle $\beta$ of H$_7$ and H$_{10}$ from $0^{\circ}$ to $45^{\circ}$, resulting in a variation from $0$ to $1$, taking $\ket{+}$ to various ending states (Fig.~\ref{fig:exx} (a)). The fidelity between the probability distributions of the unmeasured and measured cases, represented by red and blue disks respectively, are plotted against the cohering power (Fig.~\ref{fig:exx} (b)).
	The experimental data agrees very well with theoretical predictions, represented by solid lines. (Details on the calculation of $\mathcal{F}$ are provided in Supplementary Material).
	As the cohering power increases, the TPM scheme becomes more invasive, while the fidelity provided by the CM remains high. The minimal experimental observed $\mathcal{F}$ via the CM scheme is $0.963$, with a cohering power of $0.834$, while in the standard TPM approach, the minimal fidelity drops to $0.706$. The results clearly show that CM predicts transition probabilities that are closer to the unmeasured evolution.
	
	In the second part of the experiments, the above protocol is tested for a fixed $U$ with a cohering power $\sqrt{3}/2$ on input states with various initialised coherence $C_{l1}(\ket{\Phi})$ corresponding to different $p_0$ ranging from $0$ to $1$ (Fig.~\ref{fig:exy} (a)).
	The fidelity for both the CM and the TPM schemes are plotted against $p_0$  in Fig.~\ref{fig:exy} (b). In both cases, the experimental observed minimal fidelity occurs when $p_0=0.75$, with $0.906$ and $0.799$ respectively. The data matches that of theoretical fittings very well. \\

	\section{CONCLUSION}
	
	Describing work fluctuations in genuinely coherent processes remains a subtle and open question in quantum thermodynamics, although relevant progress has been achieved recently~\cite{Allahverdyan2014nonequilibrium,
		Solinas2015full,dechiara2015measuring,Talkner2016aspects,Solinas2016g,de2018ancilla,Deffner2016quantum,Miller2017time,Sampaio2017impossilbe,xu2017effects,Solinas2013work,Solinas2016probing,
		Solinas2017measurement,Hofer2017quasiprobability,Lostaglio2017quantum,NoGoTheorem,baumer2018fluctuating,PhysRevX.6.041017,PhysRevX.8.011019,miller2018leggett}.
	Here we report the first experimental observation  of work distributions, or, more precisely of transition probabilities, using an implementation based on a CM scheme~\cite{NoGoTheorem}.
	Our experimental results show how the CM scheme can reduce the measurement back-action, as compared to the standard  TPM scheme, yielding transition probabilities that are closer to the unmeasured evolution. A full understanding of the CM approach is however still in progress. For example, while relatively  elegant schemes come up in unitary  processes, similar constructions for open processes remains a challenging task.

	Our experimental results show that  quantum coherence can have an effect on the statistics, which complements previous experimental studies of work fluctuations for diagonal states~\cite{Huber2008employing,Tiago2014experimental,Shuoming2014experimental,
		Cerisola2017quantum}. Furthermore, by experimentally demonstrating the strength of the CM scheme for reducing the measurement back-action, we hope our results will stimulate new conceptual and technological developments in quantum thermodynamics and quantum information science, where CM play an important role in numerous tasks~\cite{Massar1995,bohnet2014reduced,Cox2016,hou2018deterministic}.
	\\

	\section*{METHODS}
	
	
	
	
	
	
	{\bf Details on the CM scheme}
	
	Here we provide more details on the CM scheme in Eq.~(\ref{M^{(ij)}}). Making explicit the dependence on $\lambda$, 
	\begin{align}
	M^{(ij')}_{\rm CM}(\lambda)= M^{(ij')}_{\rm TPM} \otimes \id + \lambda \ketbra{i}{i} \otimes T_{j'}^{\rm off-diag},
	\end{align}
	$\lambda$ is found by the following optimisation procedure,
	\begin{align}
	\lambda=\max_{\alpha} (\alpha : M^{(ij')}_{\rm CM}(\lambda)\geq 0 \hspace{2mm} \forall i,j).
	\end{align}
	That is, $\lambda$ is chosen so that the back action is minimised. Indeed, from Eq.~\eqref{averWCM}, it is clear that for $\lambda=1$, the back-action is minimised and the average measured work by the CM coincides with the unmeasured one in Eq.~\eqref{unmeasuredwork}. However, in general we have that $0<\lambda<1$, which ensures the positivity of the POVM elements so that this measurement scheme is operationally well defined and can be experimentally implemented.

	
	

	{\bf Details on the experimental CM}
	
	In the CM module B, the CM scheme is deterministically realized using six HWPs and four BDs, as shown in Module B of Fig.~\ref{fig:exp}. In particular, a BD displaces the horizontal-(H-) polarised photons about 3mm away from the original path while the vertical-(V-) polarised photons remmains unchanged. The action of a HWP with rotation angle $x$ is implementing a unitary transfromation on polarisation-encoded states, 
	\begin{equation}\label{hwptr}
	\begin{aligned}
	&\ket{0}\longrightarrow \cos 2x \ket{0} + \sin 2x \ket{1}, \\
	&\ket{1}\longrightarrow \sin 2x \ket{0} - \cos 2x \ket{1},
	\end{aligned}
	\end{equation}
	note that we have taken $0\equiv H$ and $1\equiv V$.

	When $\ket{\Phi}^{\otimes2}$ enters the CM module, the projector $\ketbra{i}{i}$ ($i=0,1$) in Eq.~\eqref{M^{(ij)}} on the first copy (path-encoded) is implemented as the photon enters into the $0$ or $1$ path. Then the photon goes through a two-element-POVM on the second copy (polarisation-encoded) according to the measurement outcome of the first copy. If the outcome reads $0$ (the path 1), the POVM elements on the second copy are $\cos^2\theta\mathbb{I}+\sin^2\theta\sigma_{x}$ and $2\sin^2\theta\ketbra{-}{-}$ with outcomes $00'$ and $01'$.
	
	To realise these POVMs, the rotation angle for H$_8$ is set to $67.5^{\circ}$, resulting in coherently decomposition of a pure polarisation-encoded state in the basis $\ket{+}$ and $\ket{-}$. In particular, we represent the state of Eq.~(\ref{pure1}) in the $\ket{\pm}$ basis, i.e., 
	$\ket{\Phi}=\sqrt{p_0'}\ket{+}+\sqrt{p_1'}\ket{-}$. Then from Eq.~(\ref{hwptr}), H$_8$ transforms $\ket{\Phi}$ into $\ket{\Phi'}=\sqrt{p_0'}\ket{V}+\sqrt{p_1'}\ket{H}$. Note that $\ket{\pm}=\frac{1}{\sqrt{2}}(\ket{0}\pm\ket{1})$, so $p_0=\frac{1}{2}(\sqrt{p_0'}+\sqrt{p_1'})^2$ and $p_1=\frac{1}{2}(\sqrt{p_0'}-\sqrt{p_1'})^2$. Then after passing BD$_4$, the $H$-polarised photon (aforementioned $\ket{-}$ component of $\ket{\Phi}$) is displaced by BD$_4$ and goes through a $\beta$-HWP (H$_{10}$), with a tunable angle $\beta$ controlling the parameter $\theta$ of the unitary process ($\cos^22\beta=2\sin^2\theta$). The $\beta$-HWP$_{10}$ transforms the $H$-polarised photon ($\ket{0}$) into a linearly polarised photon state $\cos2\beta\ket{0}+\sin2\beta\ket{1}$. Then BD$_7$ displaces the $\cos^22\beta$ fraction of the aforementioned $\ket{-}$ component (now $H$-polarised)  for the measurement $M^{(01')}_\theta$. The remaining $\sin^22\beta$ part of $\ket{-}$ component (now $V$-polarised) is combined with aforementioned $\ket{+}$ component of $\ket{\Phi}$ (now $H$-polarised) by BD$_5$ to obtain the measurement $M^{(00')}_\theta$. Similarly the POVM elements $M^{(10')}_\theta$ and $M^{(11')}_\theta$ can be realized by decomposing the polarisation input into $\ket{\pm}$, and letting the $\ket{+}$ component go through a H$_7$ with angle $\beta$. The two $\beta$-HWPs are highlighted in red in Fig. 1 of the main text, as this set up is capable of realizing arbitrary unitary operations $U(\theta)$, where we recall that $\cos^22\beta=2\sin^2\theta$.\\
	
	\vspace{6pt}

\appendix
	
	\section{\label{sec:appendix1}Theoretical aspects}
	\subsection{\label{sec1sub1}Basic tools for characterising coherent states and coherent evolutions}
	We start by introducing some basic tools to characterise coherent states and evolutions.
	
	\textit{Quantification of quantum coherence}. The quantification of quantum coherence starts from the definition of incoherent states, which formalises the intuition
	that quantum superpositions are non-classical. In this framework, an orthogonal basis of states $\{ \ket{k} \}$, referred to as the reference basis,  are considered as classical. Any mixture
	of such states, 
	\begin{equation}
	\chi=\sum_kp_k\ketbra{k}{k},
	\end{equation}
	with $\sum_kp_k=1$ and $p_k\geq0$, is considered classical and termed incoherent. In the frame work of quantum thermodynamics, the classical base are always taken as the eigenbasis $\{\ket{i}\}$ or $\{\ket{j'}\}$ of the Hamiltonian $H$ and $H'$. 
	
	In our experiment dealing with quantum optics, the classical base are taken as the polarisation or the path of a single photon.
	Note that in our experiments we have $i=0,1$ and $j'=0,1$.
	
	Given a quantum state $\rho$, the amount of coherence is always quantified as the distance between $\rho$ and the set of incoherent states. One of the most widely used quantification of coherence is the $l$-1 norm coherence~\cite{QuantifyingCoherence}. For a general quantum state $\rho$, it reads
	\begin{align}
	C_{l1}(\rho)=\min_{\delta\in\mathcal{I}}\|\rho-\delta\|_1,
	\label{coherence1}
	\end{align}
	where $\|A\|=\textmd{Tr}[AA^{\dag}]$ denotes the trace norm, and $\mathcal{I}$ represents the set of incoherent states. The $l$-1 norm coherence can also be expressed as
	\begin{align}
	C_{l1}(\rho)=\sum_{i\neq j}|\rho_{ij}|,
	\label{coherence2}
	\end{align}
	which is the sum of the (absolute) off-diagonal terms of $\rho$. The $l$-1 norm coherence quantifies the strength of the interference between $\ket{i}$ and $\ket{j}$. For the class of quantum state 
	\begin{equation}\label{purestate}
	\ket{\Phi}=\sqrt{p_0}\ket{0}+\sqrt{p_1}\ket{1},
	\end{equation}
	
	which is used in our experiment, the $l$-1 norm coherence simply reads 
	\begin{equation}
	C_{l1}(\ket{\Phi})=2\sqrt{p_0p_1}.
	\end{equation}
	
	
	
	\textit{Cohering power of a quantum channel}. The cohering power of a  unitary operation $U$ corresponds to the maximal coherence that can be obtained from an incoherent state by $U$.
	It can be quantified by the cohering power of a quantum channel~\cite{coheringpowerPLA},
	\begin{align}
	\mathcal{C}(\Lambda)=\max_{\delta\in\mathcal{I}}C_{l1}(\Lambda(\rho)),
	\label{cohering1}
	\end{align}
	where $\Lambda$ is a completely positive and trace preserving (CPTP) map and the optimization is taken over all incoherent states.
	For unitary processes it reads
	\begin{align}
	\mathcal{C}(U)=\|U\|^2_{l\rightarrow1}-1,
	\label{cohering2}
	\end{align}
	where $\|U\|_{l\rightarrow1}=\max\{\sum_{j}|U_{ij}|:i=1,2,3...\}$.
	Considering the unitary process
	\begin{equation}\label{coherentprocess}
	U(\theta)=\cos\theta\sigma_z+\sin\theta\sigma_x,
	\end{equation}
	which is implemented in our experiments, the cohering power can be expressed as 
	\begin{equation}
	\mathcal{C}[U(\theta)]=|\sin2\theta|.
	\end{equation}

	\subsection{\label{sec1sub2}Theoretical calculation of the fidelity}
	In this section we provide details on the calculation of the probability distribution obtained for both the CM and the TPM schemes. First, recall that these schemes are defined by the POVMs,
	\begin{align}
	M^{(ij')}_{\rm TPM}=\big|\hspace{-1mm}\bra{j'}U \ket{i}\hspace{-1mm}\big|^2 \ketbra{i}{i},
	\end{align}
	which acts on a single-copy $\rho$, and by
	\begin{align}
	M^{(ij')}_{\rm CM}= \ket{i}\bra{i} \otimes \left(\big|\hspace{-1mm}\bra{j'}U \ket{i}\hspace{-1mm}\big|^2 \id + \lambda \ketbra{i}{i} \otimes T_{j'}^{\rm off-diag}\right).
	\label{Mij}
	\end{align}
	which acts on $\rho^{\otimes 2}$, and where $T^{\rm off-diag}_{j'}$ is the off-diagonal part of
	$T_{j'}=U^{\dagger} \ketbra{j'}{j'} U$ in the $\{ \ket{i} \}$ basis.
	
	Considering the class of pure states in Eq.~(\ref{purestate}) and coherent process in Eq.~(\ref{coherentprocess}), the probability distribution of ending at state $\ket{j'} $ in the unmeasured evolution is given by
	\begin{equation}
	P^{(j')}_{\textmd{Id}}=|\bra{j'}U(\theta)\ket{\Phi}|^2, 
	\end{equation}
	yielding
	\begin{align}
	P^{(0)}_{\textmd{Id}}=(\sqrt{p_0}\cos\theta+\sqrt{p_1}\sin\theta)^2
	\end{align}
	and
	\begin{align}
	P^{(1)}_{\textmd{Id}}=(\sqrt{p_1}\cos\theta-\sqrt{p_0}\sin\theta)^2.
	\end{align}
	On the other hand, the probability distribution of the final states obtained from the CM are
	\begin{equation}
	P^{(j')}_{\textmd{CM}}=\sum_{i}P^{(ij')}_{\textmd{CM}},
	\end{equation}
	where $j'=0, 1$, and 
	\begin{equation}
	P^{(ij')}_{\rm CM}=\textmd{Tr}(M^{(ij')}_{\rm CM}\rho^{\otimes 2}). 
	\end{equation}
	By inserting $\rho=\ketbra{\Phi}{\Phi}$ and the $M^{(ij')}_{\rm CM}$ with $\lambda=\tan(\theta)$ (see main text), we obtain
	\begin{align}
	P^{(0)}_{\textmd{CM}}=p_0\cos^2\theta+p_1\sin^2\theta+2\sqrt{p_0p_1}\sin^2\theta,
	\end{align}
	and
	\begin{align}
	P^{(1)}_{\textmd{CM}}=p_0\sin^2\theta+p_1\cos^2\theta-2\sqrt{p_0p_1}\sin^2\theta.
	\end{align}
	The  fidelity between $P^{(j')}_{\textmd{CM}}$ and $P^{(j')}_{\textmd{Id}}$  reads
	\begin{equation}
	\mathcal{F}_{\textmd{CM}}=\sum_{j'}\sqrt{P^{(j')}_{\textmd{CM}}P^{(j')}_{\textmd{Id}}},
	\end{equation} 
	leading to
	\begin{equation}
	\begin{aligned}
	&\mathcal{F}_{\textmd{CM}}=|\sqrt{p_0}\cos\theta+\sqrt{p_1}\sin\theta|\times\\
	&(p_0\cos^2\theta+p_1\sin^2\theta+2\sqrt{p_0p_1}\sin^2\theta)^{\frac{1}{2}}\\
	&+|\sqrt{p_1}\cos\theta-\sqrt{p_0}\sin\theta|\times\\
	&(p_0\sin^2\theta+p_1\cos^2\theta-2\sqrt{p_0p_1}\sin^2\theta)^{\frac{1}{2}}.
	\end{aligned}
	\end{equation}
	
	Similarly, the probability distribution of the final states obtained from TPM measurements read
	\begin{equation}
	P^{(j')}_{\textmd{TPM}}=\sum_{i}P^{(ij')}_{\textmd{TPM}}, 
	\end{equation}
	where $j'=0, 1$, and 
	\begin{equation}
	P^{(ij')}_{\rm TPM}=\textmd{Tr}(M^{(ij')}_{\rm TPM}\rho). 
	\end{equation}
	Considering the state $\ket{\Phi}$ and TPM measurement 
	\begin{equation}
	M^{(ij')}_{\rm TPM}=\big|\hspace{-1mm}\bra{j'}U(\theta) \ket{i}\hspace{-1mm}\big|^2 \ketbra{i}{i}, 
	\end{equation}
	we obtain
	\begin{align}
	P^{(0)}_{\textmd{TPM}}=p_0\cos^2\theta+p_1\sin^2\theta
	\end{align}
	and
	\begin{align}
	P^{(1)}_{\textmd{TPM}}=p_0\sin^2\theta+p_1\cos^2\theta.
	\end{align}
	Then, the  fidelity between $P^{(j')}_{\textmd{TPM}}$ and $P^{(j')}_{\textmd{Id}}$ can be computed as
	\begin{equation}
	\begin{aligned}
	\label{fid2}
	&\mathcal{F}_{\textmd{CM}}=|\sqrt{p_0}\cos\theta+\sqrt{p_1}\sin\theta|(p_0\cos^2\theta+p_1\sin^2\theta)^{\frac{1}{2}}\\
	&+|\sqrt{p_1}\cos\theta-\sqrt{p_0}\sin\theta|(p_0\sin^2\theta+p_1\cos^2\theta)^{\frac{1}{2}}.
	\end{aligned}
	\end{equation}

	\section{\label{sec:appendix2}Experimental aspects}
	\subsection{\label{sec2sub1}Detail information of the single photon source}
	In module (a), a 80-mW CW laser (TopMode) with a 404-nm wavelength (linewidth=5 MHz) pumps a
	type-II beamlike phase-matching $\beta$-barium-borate (BBO, 6.0$\times$6.0$\times$2.0 $\textmd{mm}^3$, $\theta = 40.98^{\circ}$) crystal to produce a pair
	of photons with wavelength $\lambda$=808nm~\cite{takeuchi2001beamlike}. After being redirected by mirrors and passed through the interference filters (IF, $\Delta\lambda$=3nm, $\lambda$ = 808 nm), the photon pairs generated
	in spontaneous parametric down-conversion (SPDC) are coupled
	into single-mode fibers (SMF) separately. One photon (H-polarised) is used for the experiments and the other ($V$-polarised) is detected
	by a single-photon detector (SPD) acting as a trigger. The total coincidence counts are approximately $1.0\times10^4$ per second.
	
	\subsection{\label{sec2sub2}Detail information of the state preparation}
	In this section we describe the state preparation module in details. A half wave plate with rotation angle $\alpha$ can implement the unitary transition in Eq.~(\ref{coherentprocess})
	on a polarisation or path-encoded quantum state of a single photon.
	
	For generating the one-copy qubit pure state, we set the rotation angle of half-wave plate 4 (H$_4$) to $\alpha$, yielding $\ket{\Phi}$ in Eq.~(\ref{purestate}),
	where we note that $p_0=|\cos2\alpha|^2$ and $p_1=|\sin2\alpha|^2$. Thus the desired qubit state is prepared.

	For experimentally generating identically two pure qubit product state, we take advantage of the multiple degrees of a single photon, encoded in polarisation and path. Initially, a single-photon $\ket{H}$ is generated. It passes through H$_1$ with a rotation angle $\alpha$, resulting in a pure polarisation-encoded qubit state in Eq.~(\ref{purestate}). Then, the photon passes the beam displacer 1 (BD$_1$), the $H$ component is displaced into path $0$, which is 4 mm away from the $V$ component in path $1$, resulting in a path-polarised entangled state
	\begin{equation}
	\sqrt{p_0}\ket{0}\otimes\ket{H}+\sqrt{p_1}\ket{1}\otimes\ket{V}.
	\end{equation}
	Following, H$_2$ with a rotation angle of $45^{\circ}$ flips the $V$-polarised photon to a $H$-polarised photon, resulting in a product state 
	\begin{equation}
	\ket{\Phi}\otimes\ket{H}, 
	\end{equation}
	where $\ket{\Phi}$ denotes state in Eq.~(\ref{purestate}). Finally,  the second copy is encoded into the polarisation degree by setting H$_3$ to $\alpha$, thus generating the desired state $\ket{\Phi}^{\otimes2}$. In our experiments, we use the convention $\{ \ket{0},\ket{1}\}\equiv \{\ket{H},\ket{V}\}$.
	
	\subsection{\label{sec2sub3}Experimental data}
	Experimental data for two experiments are list in Table S1 and S2. Here, $\tilde{\mathcal{F}}_{\textmd{CM}}$ ($\tilde{\mathcal{F}}_{\textmd{TPM}}$) denotes the fidelity between unmeasured final states and the transition probability obtained from CM scheme (TPM scheme). For the experiment conducted with different process and a fixed maximally coherent input $\ket{+}=(\ket{0}+\ket{1})/\sqrt{2}$, we use the class of unitary process Eq.~(\ref{coherentprocess}). Results are tabulated for various configured
	values of $\beta$ and cohering power $\mathcal{C}$, where we recall that $\cos^22\beta=2\sin^2\theta$ and $\mathcal{C}[U(\theta)]=|\sin2\theta|$. For the experiment conducted with pure states with various initialized coherence and a fixed coherent process $U(\pi/6)$, we experimentally prepare $\ket{\Phi}$ in Eq.~(\ref{purestate}). Results are tabulated for the rotation angle $\alpha$ of H$_{1,2,3,4}$, where $\cos^22\alpha=p_0$ and $\sin^22\alpha=p_0$.
	
	\begin{table}[htp!]\label{tab:process}
		\begin{tabular}{c|cccccccc}
			\hline
			\hline
			$\beta$($\circ$)& 0& 3& 6& 9& 12& 15 & 18& 21\\
			\hline
			$\theta$($\circ$)&0.0& 4.24& 8.454& 12.6& 16.7& 20.7& 24.6& 28.2\\
			\hline
			$\mathcal{C}$&0.0& 0.147& 0.291& 0.426& 0.551& 0.661& 0.756& 0.834\\
			\hline
			$\tilde{\mathcal{F}}_{\textmd{CM}}$&1.00&0.997&0.991&0.982&0.974&0.975&0.966&0.963\\
			\hline
			$\tilde{\mathcal{F}}_{\textmd{TPM}}$&1.00&0.997&0.990&0.978&0.960&0.938&0.910&0.883\\
			\hline
			\hline
			$\beta$($\circ$)& 24& 27& 30& 33& 36& 39& 42& 45\\
			\hline
			$\theta$($\circ$)& 31.7& 34.9& 37.8& 40.2& 42.3& 43.8& 44.7& 45.0\\
			\hline
			$\mathcal{C}$& 0.894& 0.938& 0.968& 0.986& 0.995& 0.999& 1.00& 1.00\\
			\hline
			$\tilde{\mathcal{F}}_{\textmd{CM}}$&0.968&0.968&0.974&0.977&0.984&0.991&0.997&0.998\\
			\hline
			$\tilde{\mathcal{F}}_{\textmd{TPM}}$&0.850&0.822&0.790&0.740&0.732&0.720&0.708&0.706\\
			\hline
			\hline
		\end{tabular}
		\caption{\label{tab:process} Experimental data for different coherent process}
	\end{table}
	
	\begin{table}[htp!]\label{tab:state}
		\begin{tabular}{c|cccccccc}
			\hline
			\hline
			$\alpha$($\circ$)& 0& 3& 6& 9& 12& 15 & 18& 21\\
			\hline
			$p_0$&1.00& 0.989& 0.957& 0.905& 0.835& 0.75& 0.655& 0.552\\
			\hline
			$\tilde{\mathcal{F}}_{\textmd{CM}}$&1.00&1.00&0.995&0.977&0.955&0.906&0.927&0.949\\
			\hline
			$\tilde{\mathcal{F}}_{\textmd{TPM}}$&1.00&0.996&0.976&0.936&0.883&0.799&0.830&0.860\\
			\hline
			\hline
			$\alpha$($\circ$)& 24& 27& 30& 33& 36& 39& 42& 45\\
			\hline
			$p_0$& 0.448& 0.345& 0.25& 0.165& 0.095& 0.043& 0.011& 0.00\\
			\hline
			
			$\tilde{\mathcal{F}}_{\textmd{CM}}$&0.963&0.973&0.982&0.989&0.994&0.997&0.999&1.00\\
			\hline
			$\tilde{\mathcal{F}}_{\textmd{TPM}}$&0.883&0.900&0.921&0.943&0.963&0.979&0.992&1.00\\
			\hline
			\hline
		\end{tabular}
		\caption{\label{tab:state} Experimental data for states with various initial coherence}
	\end{table}
	
	Both experiments are carried out repeatedly for recording proportion coincidence counts
	\begin{align}
	\tilde{P}^{(ij')}_{\textmd{CM}}=\frac{\tilde{N}^{(ij')}_{\textmd{CM}}}{\sum_{i,j}\tilde{N}^{(ij')}_{\textmd{CM}}}.
	\label{exppro1}
	\end{align}
	and
	\begin{align}
	\tilde{P}^{(ij')}_{\textmd{TPM}}=\frac{\tilde{N}^{(ij')}_{\textmd{TPM}}}{\sum_{i,j}\tilde{N}^{(ij')}_{\textmd{TPM}}}.
	\label{exppro2}
	\end{align}
	where $\tilde{N}^{(ij')}_{\textmd{CM}}$ ($\tilde{N}^{(ij')}_{\textmd{TPM}}$) denotes the coincident events associated with each POVM out comes $ij'$.
	
	The experimental fidelity between the ideal transition and two measurement schemes are obtained as
	\begin{align}
	\tilde{\mathcal{F}}_{\textmd{CM}}=\sqrt{\tilde{P}^{(0')}_{\textmd{CM}}P^{(0')}_{\textmd{Id}}}+\sqrt{\tilde{P}^{(1')}_{\textmd{CM}}P^{(1')}_{\textmd{Id}}}.
	\label{expfid1}
	\end{align}
	and
	\begin{align}
	\tilde{\mathcal{F}}_{\textmd{CM}}=\sqrt{\tilde{P}^{(0')}_{\textmd{TPM}}P^{(0')}_{\textmd{Id}}}+\sqrt{\tilde{P}^{(1')}_{\textmd{TPM}}P^{(1')}_{\textmd{Id}}}.
	\label{expfid2}
	\end{align}
	where $\tilde{P}^{(j')}_{\textmd{CM}}=\sum_{i}\tilde{P}^{(ij')}_{\textmd{CM}}$ and $\tilde{P}^{(j')}_{\textmd{TPM}}=\sum_{i}\tilde{P}^{(ij')}_{\textmd{TPM}}$ are experimental data obtained from Eq.~(\ref{exppro1}) and Eq.~(\ref{exppro2}).


\begin{thebibliography}{44}%
	\makeatletter
	\providecommand \@ifxundefined [1]{%
		\@ifx{#1\undefined}
	}%
	\providecommand \@ifnum [1]{%
		\ifnum #1\expandafter \@firstoftwo
		\else \expandafter \@secondoftwo
		\fi
	}%
	\providecommand \@ifx [1]{%
		\ifx #1\expandafter \@firstoftwo
		\else \expandafter \@secondoftwo
		\fi
	}%
	\providecommand \natexlab [1]{#1}%
	\providecommand \enquote  [1]{``#1''}%
	\providecommand \bibnamefont  [1]{#1}%
	\providecommand \bibfnamefont [1]{#1}%
	\providecommand \citenamefont [1]{#1}%
	\providecommand \href@noop [0]{\@secondoftwo}%
	\providecommand \href [0]{\begingroup \@sanitize@url \@href}%
	\providecommand \@href[1]{\@@startlink{#1}\@@href}%
	\providecommand \@@href[1]{\endgroup#1\@@endlink}%
	\providecommand \@sanitize@url [0]{\catcode `\\12\catcode `\$12\catcode
		`\&12\catcode `\#12\catcode `\^12\catcode `\_12\catcode `\%12\relax}%
	\providecommand \@@startlink[1]{}%
	\providecommand \@@endlink[0]{}%
	\providecommand \url  [0]{\begingroup\@sanitize@url \@url }%
	\providecommand \@url [1]{\endgroup\@href {#1}{\urlprefix }}%
	\providecommand \urlprefix  [0]{URL }%
	\providecommand \Eprint [0]{\href }%
	\providecommand \doibase [0]{http://dx.doi.org/}%
	\providecommand \selectlanguage [0]{\@gobble}%
	\providecommand \bibinfo  [0]{\@secondoftwo}%
	\providecommand \bibfield  [0]{\@secondoftwo}%
	\providecommand \translation [1]{[#1]}%
	\providecommand \BibitemOpen [0]{}%
	\providecommand \bibitemStop [0]{}%
	\providecommand \bibitemNoStop [0]{.\EOS\space}%
	\providecommand \EOS [0]{\spacefactor3000\relax}%
	\providecommand \BibitemShut  [1]{\csname bibitem#1\endcsname}%
	\let\auto@bib@innerbib\@empty
	\bibitem [{\citenamefont {Campisi}\ \emph {et~al.}(2011)\citenamefont
		{Campisi}, \citenamefont {H\"anggi},\ and\ \citenamefont
		{Talkner}}]{TPM1RMP}%
	\BibitemOpen
	\bibfield  {author} {\bibinfo {author} {\bibfnamefont {M.}~\bibnamefont
			{Campisi}}, \bibinfo {author} {\bibfnamefont {P.}~\bibnamefont {H\"anggi}}, \
		and\ \bibinfo {author} {\bibfnamefont {P.}~\bibnamefont {Talkner}},\ }\href
	{\doibase 10.1103/RevModPhys.83.771} {\bibfield  {journal} {\bibinfo
			{journal} {Rev. Mod. Phys.}\ }\textbf {\bibinfo {volume} {83}},\ \bibinfo
		{pages} {771} (\bibinfo {year} {2011})}\BibitemShut {NoStop}%
	\bibitem [{\citenamefont {Talkner}\ \emph {et~al.}(2007)\citenamefont
		{Talkner}, \citenamefont {Lutz},\ and\ \citenamefont {H\"anggi}}]{TPM2PRE}%
	\BibitemOpen
	\bibfield  {author} {\bibinfo {author} {\bibfnamefont {P.}~\bibnamefont
			{Talkner}}, \bibinfo {author} {\bibfnamefont {E.}~\bibnamefont {Lutz}}, \
		and\ \bibinfo {author} {\bibfnamefont {P.}~\bibnamefont {H\"anggi}},\ }\href
	{\doibase 10.1103/PhysRevE.75.050102} {\bibfield  {journal} {\bibinfo
			{journal} {Phys. Rev. E}\ }\textbf {\bibinfo {volume} {75}},\ \bibinfo
		{pages} {050102} (\bibinfo {year} {2007})}\BibitemShut {NoStop}%
	\bibitem [{\citenamefont
		{Allahverdyan}(2014)}]{Allahverdyan2014nonequilibrium}%
	\BibitemOpen
	\bibfield  {author} {\bibinfo {author} {\bibfnamefont {A.~E.}\ \bibnamefont
			{Allahverdyan}},\ }\href {\doibase 10.1103/PhysRevE.90.032137} {\bibfield
		{journal} {\bibinfo  {journal} {Phys. Rev. E}\ }\textbf {\bibinfo {volume}
			{90}},\ \bibinfo {pages} {032137} (\bibinfo {year} {2014})}\BibitemShut
	{NoStop}%
	\bibitem [{\citenamefont {Solinas}\ and\ \citenamefont
		{Gasparinetti}(2015)}]{Solinas2015full}%
	\BibitemOpen
	\bibfield  {author} {\bibinfo {author} {\bibfnamefont {P.}~\bibnamefont
			{Solinas}}\ and\ \bibinfo {author} {\bibfnamefont {S.}~\bibnamefont
			{Gasparinetti}},\ }\href {\doibase 10.1103/PhysRevE.92.042150} {\bibfield
		{journal} {\bibinfo  {journal} {Phys. Rev. E}\ }\textbf {\bibinfo {volume}
			{92}},\ \bibinfo {pages} {042150} (\bibinfo {year} {2015})}\BibitemShut
	{NoStop}%
	\bibitem [{\citenamefont {Chiara}\ \emph {et~al.}(2015)\citenamefont {Chiara},
		\citenamefont {Roncaglia},\ and\ \citenamefont
		{Paz}}]{dechiara2015measuring}%
	\BibitemOpen
	\bibfield  {author} {\bibinfo {author} {\bibfnamefont {G.~D.}\ \bibnamefont
			{Chiara}}, \bibinfo {author} {\bibfnamefont {A.~J.}\ \bibnamefont
			{Roncaglia}}, \ and\ \bibinfo {author} {\bibfnamefont {J.~P.}\ \bibnamefont
			{Paz}},\ }\href {http://stacks.iop.org/1367-2630/17/i=3/a=035004} {\bibfield
		{journal} {\bibinfo  {journal} {New Journal of Physics}\ }\textbf {\bibinfo
			{volume} {17}},\ \bibinfo {pages} {035004} (\bibinfo {year}
		{2015})}\BibitemShut {NoStop}%
	\bibitem [{\citenamefont {Talkner}\ and\ \citenamefont
		{H\"anggi}(2016)}]{Talkner2016aspects}%
	\BibitemOpen
	\bibfield  {author} {\bibinfo {author} {\bibfnamefont {P.}~\bibnamefont
			{Talkner}}\ and\ \bibinfo {author} {\bibfnamefont {P.}~\bibnamefont
			{H\"anggi}},\ }\href {\doibase 10.1103/PhysRevE.93.022131} {\bibfield
		{journal} {\bibinfo  {journal} {Phys. Rev. E}\ }\textbf {\bibinfo {volume}
			{93}},\ \bibinfo {pages} {022131} (\bibinfo {year} {2016})}\BibitemShut
	{NoStop}%
	\bibitem [{\citenamefont {Solinas}\ and\ \citenamefont
		{Gasparinetti}(2016{\natexlab{a}})}]{Solinas2016g}%
	\BibitemOpen
	\bibfield  {author} {\bibinfo {author} {\bibfnamefont {P.}~\bibnamefont
			{Solinas}}\ and\ \bibinfo {author} {\bibfnamefont {S.}~\bibnamefont
			{Gasparinetti}},\ }\href {\doibase 10.1103/PhysRevA.94.052103} {\bibfield
		{journal} {\bibinfo  {journal} {Phys. Rev. A}\ }\textbf {\bibinfo {volume}
			{94}},\ \bibinfo {pages} {052103} (\bibinfo {year}
		{2016}{\natexlab{a}})}\BibitemShut {NoStop}%
	\bibitem [{\citenamefont {De~Chiara}\ \emph {et~al.}(2018)\citenamefont
		{De~Chiara}, \citenamefont {Solinas}, \citenamefont {Cerisola},\ and\
		\citenamefont {Roncaglia}}]{de2018ancilla}%
	\BibitemOpen
	\bibfield  {author} {\bibinfo {author} {\bibfnamefont {G.}~\bibnamefont
			{De~Chiara}}, \bibinfo {author} {\bibfnamefont {P.}~\bibnamefont {Solinas}},
		\bibinfo {author} {\bibfnamefont {F.}~\bibnamefont {Cerisola}}, \ and\
		\bibinfo {author} {\bibfnamefont {A.~J.}\ \bibnamefont {Roncaglia}},\
	}\href@noop {} {\bibfield  {journal} {\bibinfo  {journal} {arXiv preprint
				arXiv:1805.06047}\ } (\bibinfo {year} {2018})}\BibitemShut {NoStop}%
	\bibitem [{\citenamefont {Deffner}\ \emph {et~al.}(2016)\citenamefont
		{Deffner}, \citenamefont {Paz},\ and\ \citenamefont
		{Zurek}}]{Deffner2016quantum}%
	\BibitemOpen
	\bibfield  {author} {\bibinfo {author} {\bibfnamefont {S.}~\bibnamefont
			{Deffner}}, \bibinfo {author} {\bibfnamefont {J.~P.}\ \bibnamefont {Paz}}, \
		and\ \bibinfo {author} {\bibfnamefont {W.~H.}\ \bibnamefont {Zurek}},\ }\href
	{\doibase 10.1103/PhysRevE.94.010103} {\bibfield  {journal} {\bibinfo
			{journal} {Phys. Rev. E}\ }\textbf {\bibinfo {volume} {94}},\ \bibinfo
		{pages} {010103} (\bibinfo {year} {2016})}\BibitemShut {NoStop}%
	\bibitem [{\citenamefont {Miller}\ and\ \citenamefont
		{Anders}(2017)}]{Miller2017time}%
	\BibitemOpen
	\bibfield  {author} {\bibinfo {author} {\bibfnamefont {H.~J.~D.}\
			\bibnamefont {Miller}}\ and\ \bibinfo {author} {\bibfnamefont
			{J.}~\bibnamefont {Anders}},\ }\href
	{http://stacks.iop.org/1367-2630/19/i=6/a=062001} {\bibfield  {journal}
		{\bibinfo  {journal} {New Journal of Physics}\ }\textbf {\bibinfo {volume}
			{19}},\ \bibinfo {pages} {062001} (\bibinfo {year} {2017})}\BibitemShut
	{NoStop}%
	\bibitem [{\citenamefont {Sampaio}\ \emph {et~al.}(2018)\citenamefont
		{Sampaio}, \citenamefont {Suomela}, \citenamefont {Ala-Nissila},
		\citenamefont {Anders},\ and\ \citenamefont
		{Philbin}}]{Sampaio2017impossilbe}%
	\BibitemOpen
	\bibfield  {author} {\bibinfo {author} {\bibfnamefont {R.}~\bibnamefont
			{Sampaio}}, \bibinfo {author} {\bibfnamefont {S.}~\bibnamefont {Suomela}},
		\bibinfo {author} {\bibfnamefont {T.}~\bibnamefont {Ala-Nissila}}, \bibinfo
		{author} {\bibfnamefont {J.}~\bibnamefont {Anders}}, \ and\ \bibinfo {author}
		{\bibfnamefont {T.~G.}\ \bibnamefont {Philbin}},\ }\href {\doibase
		10.1103/PhysRevA.97.012131} {\bibfield  {journal} {\bibinfo  {journal} {Phys.
				Rev. A}\ }\textbf {\bibinfo {volume} {97}},\ \bibinfo {pages} {012131}
		(\bibinfo {year} {2018})}\BibitemShut {NoStop}%
	\bibitem [{\citenamefont {Xu}\ \emph {et~al.}(2018)\citenamefont {Xu},
		\citenamefont {Zou}, \citenamefont {Guo},\ and\ \citenamefont
		{Kong}}]{xu2017effects}%
	\BibitemOpen
	\bibfield  {author} {\bibinfo {author} {\bibfnamefont {B.-M.}\ \bibnamefont
			{Xu}}, \bibinfo {author} {\bibfnamefont {J.}~\bibnamefont {Zou}}, \bibinfo
		{author} {\bibfnamefont {L.-S.}\ \bibnamefont {Guo}}, \ and\ \bibinfo
		{author} {\bibfnamefont {X.-M.}\ \bibnamefont {Kong}},\ }\href {\doibase
		10.1103/PhysRevA.97.052122} {\bibfield  {journal} {\bibinfo  {journal} {Phys.
				Rev. A}\ }\textbf {\bibinfo {volume} {97}},\ \bibinfo {pages} {052122}
		(\bibinfo {year} {2018})}\BibitemShut {NoStop}%
	\bibitem [{\citenamefont {Solinas}\ \emph {et~al.}(2013)\citenamefont
		{Solinas}, \citenamefont {Averin},\ and\ \citenamefont
		{Pekola}}]{Solinas2013work}%
	\BibitemOpen
	\bibfield  {author} {\bibinfo {author} {\bibfnamefont {P.}~\bibnamefont
			{Solinas}}, \bibinfo {author} {\bibfnamefont {D.~V.}\ \bibnamefont {Averin}},
		\ and\ \bibinfo {author} {\bibfnamefont {J.~P.}\ \bibnamefont {Pekola}},\
	}\href {\doibase 10.1103/PhysRevB.87.060508} {\bibfield  {journal} {\bibinfo
			{journal} {Phys. Rev. B}\ }\textbf {\bibinfo {volume} {87}},\ \bibinfo
		{pages} {060508} (\bibinfo {year} {2013})}\BibitemShut {NoStop}%
	\bibitem [{\citenamefont {Solinas}\ and\ \citenamefont
		{Gasparinetti}(2016{\natexlab{b}})}]{Solinas2016probing}%
	\BibitemOpen
	\bibfield  {author} {\bibinfo {author} {\bibfnamefont {P.}~\bibnamefont
			{Solinas}}\ and\ \bibinfo {author} {\bibfnamefont {S.}~\bibnamefont
			{Gasparinetti}},\ }\href {\doibase 10.1103/PhysRevA.94.052103} {\bibfield
		{journal} {\bibinfo  {journal} {Phys. Rev. A}\ }\textbf {\bibinfo {volume}
			{94}},\ \bibinfo {pages} {052103} (\bibinfo {year}
		{2016}{\natexlab{b}})}\BibitemShut {NoStop}%
	\bibitem [{\citenamefont {Solinas}\ \emph {et~al.}(2017)\citenamefont
		{Solinas}, \citenamefont {Miller},\ and\ \citenamefont
		{Anders}}]{Solinas2017measurement}%
	\BibitemOpen
	\bibfield  {author} {\bibinfo {author} {\bibfnamefont {P.}~\bibnamefont
			{Solinas}}, \bibinfo {author} {\bibfnamefont {H.~J.~D.}\ \bibnamefont
			{Miller}}, \ and\ \bibinfo {author} {\bibfnamefont {J.}~\bibnamefont
			{Anders}},\ }\href {\doibase 10.1103/PhysRevA.96.052115} {\bibfield
		{journal} {\bibinfo  {journal} {Phys. Rev. A}\ }\textbf {\bibinfo {volume}
			{96}},\ \bibinfo {pages} {052115} (\bibinfo {year} {2017})}\BibitemShut
	{NoStop}%
	\bibitem [{\citenamefont {Hofer}(2017)}]{Hofer2017quasiprobability}%
	\BibitemOpen
	\bibfield  {author} {\bibinfo {author} {\bibfnamefont {P.~P.}\ \bibnamefont
			{Hofer}},\ }\href {\doibase 10.22331/q-2017-10-12-32} {\bibfield  {journal}
		{\bibinfo  {journal} {{Quantum}}\ }\textbf {\bibinfo {volume} {1}},\ \bibinfo
		{pages} {32} (\bibinfo {year} {2017})}\BibitemShut {NoStop}%
	\bibitem [{\citenamefont {Lostaglio}(2018)}]{Lostaglio2017quantum}%
	\BibitemOpen
	\bibfield  {author} {\bibinfo {author} {\bibfnamefont {M.}~\bibnamefont
			{Lostaglio}},\ }\href {\doibase 10.1103/PhysRevLett.120.040602} {\bibfield
		{journal} {\bibinfo  {journal} {Phys. Rev. Lett.}\ }\textbf {\bibinfo
			{volume} {120}},\ \bibinfo {pages} {040602} (\bibinfo {year}
		{2018})}\BibitemShut {NoStop}%
	\bibitem [{\citenamefont {Perarnau-Llobet}\ \emph {et~al.}(2017)\citenamefont
		{Perarnau-Llobet}, \citenamefont {B\"aumer}, \citenamefont {Hovhannisyan},
		\citenamefont {Huber},\ and\ \citenamefont {Acin}}]{NoGoTheorem}%
	\BibitemOpen
	\bibfield  {author} {\bibinfo {author} {\bibfnamefont {M.}~\bibnamefont
			{Perarnau-Llobet}}, \bibinfo {author} {\bibfnamefont {E.}~\bibnamefont
			{B\"aumer}}, \bibinfo {author} {\bibfnamefont {K.~V.}\ \bibnamefont
			{Hovhannisyan}}, \bibinfo {author} {\bibfnamefont {M.}~\bibnamefont {Huber}},
		\ and\ \bibinfo {author} {\bibfnamefont {A.}~\bibnamefont {Acin}},\ }\href
	{\doibase 10.1103/PhysRevLett.118.070601} {\bibfield  {journal} {\bibinfo
			{journal} {Phys. Rev. Lett.}\ }\textbf {\bibinfo {volume} {118}},\ \bibinfo
		{pages} {070601} (\bibinfo {year} {2017})}\BibitemShut {NoStop}%
	\bibitem [{\citenamefont {B{\"a}umer}\ \emph {et~al.}(2018)\citenamefont
		{B{\"a}umer}, \citenamefont {Lostaglio}, \citenamefont {Perarnau-Llobet},\
		and\ \citenamefont {Sampaio}}]{baumer2018fluctuating}%
	\BibitemOpen
	\bibfield  {author} {\bibinfo {author} {\bibfnamefont {E.}~\bibnamefont
			{B{\"a}umer}}, \bibinfo {author} {\bibfnamefont {M.}~\bibnamefont
			{Lostaglio}}, \bibinfo {author} {\bibfnamefont {M.}~\bibnamefont
			{Perarnau-Llobet}}, \ and\ \bibinfo {author} {\bibfnamefont {R.}~\bibnamefont
			{Sampaio}},\ }\href {https://arxiv.org/abs/1805.10096} {\bibfield  {journal}
		{\bibinfo  {journal} {arXiv preprint arXiv:1805.10096}\ } (\bibinfo {year}
		{2018})}\BibitemShut {NoStop}%
	\bibitem [{\citenamefont {Korzekwa}\ \emph {et~al.}(2016)\citenamefont
		{Korzekwa}, \citenamefont {Lostaglio}, \citenamefont {Oppenheim},\ and\
		\citenamefont {Jennings}}]{korzekwa2016extraction}%
	\BibitemOpen
	\bibfield  {author} {\bibinfo {author} {\bibfnamefont {K.}~\bibnamefont
			{Korzekwa}}, \bibinfo {author} {\bibfnamefont {M.}~\bibnamefont {Lostaglio}},
		\bibinfo {author} {\bibfnamefont {J.}~\bibnamefont {Oppenheim}}, \ and\
		\bibinfo {author} {\bibfnamefont {D.}~\bibnamefont {Jennings}},\ }\href@noop
	{} {\bibfield  {journal} {\bibinfo  {journal} {New Journal of Physics}\
		}\textbf {\bibinfo {volume} {18}},\ \bibinfo {pages} {023045} (\bibinfo
		{year} {2016})}\BibitemShut {NoStop}%
	\bibitem [{\citenamefont {L{\"o}rch}\ \emph {et~al.}(2018)\citenamefont
		{L{\"o}rch}, \citenamefont {Bruder}, \citenamefont {Brunner},\ and\
		\citenamefont {Hofer}}]{Patrick2018}%
	\BibitemOpen
	\bibfield  {author} {\bibinfo {author} {\bibfnamefont {N.}~\bibnamefont
			{L{\"o}rch}}, \bibinfo {author} {\bibfnamefont {C.}~\bibnamefont {Bruder}},
		\bibinfo {author} {\bibfnamefont {N.}~\bibnamefont {Brunner}}, \ and\
		\bibinfo {author} {\bibfnamefont {P.~P.}\ \bibnamefont {Hofer}},\ }\href
	{http://iopscience.iop.org/article/10.1088/2058-9565/aacbf3/meta} {\bibfield
		{journal} {\bibinfo  {journal} {Quantum Science and Technology}\ } (\bibinfo
		{year} {2018})}\BibitemShut {NoStop}%
	\bibitem [{\citenamefont {Ro\ss{}nagel}\ \emph {et~al.}(2014)\citenamefont
		{Ro\ss{}nagel}, \citenamefont {Abah}, \citenamefont {Schmidt-Kaler},
		\citenamefont {Singer},\ and\ \citenamefont {Lutz}}]{Rossnagel2014nanoscale}%
	\BibitemOpen
	\bibfield  {author} {\bibinfo {author} {\bibfnamefont {J.}~\bibnamefont
			{Ro\ss{}nagel}}, \bibinfo {author} {\bibfnamefont {O.}~\bibnamefont {Abah}},
		\bibinfo {author} {\bibfnamefont {F.}~\bibnamefont {Schmidt-Kaler}}, \bibinfo
		{author} {\bibfnamefont {K.}~\bibnamefont {Singer}}, \ and\ \bibinfo {author}
		{\bibfnamefont {E.}~\bibnamefont {Lutz}},\ }\href {\doibase
		10.1103/PhysRevLett.112.030602} {\bibfield  {journal} {\bibinfo  {journal}
			{Phys. Rev. Lett.}\ }\textbf {\bibinfo {volume} {112}},\ \bibinfo {pages}
		{030602} (\bibinfo {year} {2014})}\BibitemShut {NoStop}%
	\bibitem [{\citenamefont {Correa}\ \emph {et~al.}(2014)\citenamefont {Correa},
		\citenamefont {Palao}, \citenamefont {Alonso},\ and\ \citenamefont
		{Adesso}}]{Correa2014quantum}%
	\BibitemOpen
	\bibfield  {author} {\bibinfo {author} {\bibfnamefont {L.~A.}\ \bibnamefont
			{Correa}}, \bibinfo {author} {\bibfnamefont {J.~P.}\ \bibnamefont {Palao}},
		\bibinfo {author} {\bibfnamefont {D.}~\bibnamefont {Alonso}}, \ and\ \bibinfo
		{author} {\bibfnamefont {G.}~\bibnamefont {Adesso}},\ }\href {\doibase
		10.1038/srep03949} {\bibfield  {journal} {\bibinfo  {journal} {Scientific
				Reports}\ }\textbf {\bibinfo {volume} {4}},\ \bibinfo {pages} {3949}
		(\bibinfo {year} {2014})}\BibitemShut {NoStop}%
	\bibitem [{\citenamefont {Mitchison}\ \emph {et~al.}(2015)\citenamefont
		{Mitchison}, \citenamefont {Woods}, \citenamefont {Prior},\ and\
		\citenamefont {Huber}}]{Mitchison2015Coherence}%
	\BibitemOpen
	\bibfield  {author} {\bibinfo {author} {\bibfnamefont {M.~T.}\ \bibnamefont
			{Mitchison}}, \bibinfo {author} {\bibfnamefont {M.~P.}\ \bibnamefont
			{Woods}}, \bibinfo {author} {\bibfnamefont {J.}~\bibnamefont {Prior}}, \ and\
		\bibinfo {author} {\bibfnamefont {M.}~\bibnamefont {Huber}},\ }\href
	{http://stacks.iop.org/1367-2630/17/i=11/a=115013} {\bibfield  {journal}
		{\bibinfo  {journal} {New Journal of Physics}\ }\textbf {\bibinfo {volume}
			{17}},\ \bibinfo {pages} {115013} (\bibinfo {year} {2015})}\BibitemShut
	{NoStop}%
	\bibitem [{\citenamefont {Klatzow}\ \emph {et~al.}(2017)\citenamefont
		{Klatzow}, \citenamefont {Becker}, \citenamefont {Ledingham}, \citenamefont
		{Weinzetl}, \citenamefont {Kaczmarek}, \citenamefont {Saunders},
		\citenamefont {Nunn}, \citenamefont {Walmsley}, \citenamefont {Uzdin},\ and\
		\citenamefont {Poem}}]{Klatzow2017Experimental}%
	\BibitemOpen
	\bibfield  {author} {\bibinfo {author} {\bibfnamefont {J.}~\bibnamefont
			{Klatzow}}, \bibinfo {author} {\bibfnamefont {J.~N.}\ \bibnamefont {Becker}},
		\bibinfo {author} {\bibfnamefont {P.~M.}\ \bibnamefont {Ledingham}}, \bibinfo
		{author} {\bibfnamefont {C.}~\bibnamefont {Weinzetl}}, \bibinfo {author}
		{\bibfnamefont {K.~T.}\ \bibnamefont {Kaczmarek}}, \bibinfo {author}
		{\bibfnamefont {D.~J.}\ \bibnamefont {Saunders}}, \bibinfo {author}
		{\bibfnamefont {J.}~\bibnamefont {Nunn}}, \bibinfo {author} {\bibfnamefont
			{I.~A.}\ \bibnamefont {Walmsley}}, \bibinfo {author} {\bibfnamefont
			{R.}~\bibnamefont {Uzdin}}, \ and\ \bibinfo {author} {\bibfnamefont
			{E.}~\bibnamefont {Poem}},\ }\href@noop {} {\bibfield  {journal} {\bibinfo
			{journal} {arXiv preprint arXiv:1710.08716}\ } (\bibinfo {year}
		{2017})}\BibitemShut {NoStop}%
	\bibitem [{\citenamefont {Lostaglio}\ \emph {et~al.}(2015)\citenamefont
		{Lostaglio}, \citenamefont {Jennings},\ and\ \citenamefont
		{Rudolph}}]{lostaglio2015description}%
	\BibitemOpen
	\bibfield  {author} {\bibinfo {author} {\bibfnamefont {M.}~\bibnamefont
			{Lostaglio}}, \bibinfo {author} {\bibfnamefont {D.}~\bibnamefont {Jennings}},
		\ and\ \bibinfo {author} {\bibfnamefont {T.}~\bibnamefont {Rudolph}},\
	}\href@noop {} {\bibfield  {journal} {\bibinfo  {journal} {Nature
				communications}\ }\textbf {\bibinfo {volume} {6}} (\bibinfo {year}
		{2015})}\BibitemShut {NoStop}%
	\bibitem [{\citenamefont {Janzing}(2006)}]{Janzing2006}%
	\BibitemOpen
	\bibfield  {author} {\bibinfo {author} {\bibfnamefont {D.}~\bibnamefont
			{Janzing}},\ }\href {\doibase 10.1007/s10955-006-9220-x} {\bibfield
		{journal} {\bibinfo  {journal} {Journal of Statistical Physics}\ }\textbf
		{\bibinfo {volume} {125}},\ \bibinfo {pages} {761} (\bibinfo {year}
		{2006})}\BibitemShut {NoStop}%
	\bibitem [{\citenamefont {Giovannetti}\ \emph {et~al.}(2011)\citenamefont
		{Giovannetti}, \citenamefont {Lloyd},\ and\ \citenamefont
		{Maccone}}]{Giov11advances}%
	\BibitemOpen
	\bibfield  {author} {\bibinfo {author} {\bibfnamefont {V.}~\bibnamefont
			{Giovannetti}}, \bibinfo {author} {\bibfnamefont {S.}~\bibnamefont {Lloyd}},
		\ and\ \bibinfo {author} {\bibfnamefont {L.}~\bibnamefont {Maccone}},\
	}\href@noop {} {\bibfield  {journal} {\bibinfo  {journal} {Nat. Photon.}\
		}\textbf {\bibinfo {volume} {5}},\ \bibinfo {pages} {222} (\bibinfo {year}
		{2011})}\BibitemShut {NoStop}%
	\bibitem [{\citenamefont {Bohnet}\ \emph {et~al.}(2014)\citenamefont {Bohnet},
		\citenamefont {Cox}, \citenamefont {Norcia}, \citenamefont {Weiner},
		\citenamefont {Chen},\ and\ \citenamefont {Thompson}}]{bohnet2014reduced}%
	\BibitemOpen
	\bibfield  {author} {\bibinfo {author} {\bibfnamefont {J.~G.}\ \bibnamefont
			{Bohnet}}, \bibinfo {author} {\bibfnamefont {K.~C.}\ \bibnamefont {Cox}},
		\bibinfo {author} {\bibfnamefont {M.~A.}\ \bibnamefont {Norcia}}, \bibinfo
		{author} {\bibfnamefont {J.~M.}\ \bibnamefont {Weiner}}, \bibinfo {author}
		{\bibfnamefont {Z.}~\bibnamefont {Chen}}, \ and\ \bibinfo {author}
		{\bibfnamefont {J.~K.}\ \bibnamefont {Thompson}},\ }\href@noop {} {\bibfield
		{journal} {\bibinfo  {journal} {Nature Photonics}\ }\textbf {\bibinfo
			{volume} {8}},\ \bibinfo {pages} {731} (\bibinfo {year} {2014})}\BibitemShut
	{NoStop}%
	\bibitem [{\citenamefont {Massar}\ and\ \citenamefont
		{Popescu}(1995)}]{Massar1995}%
	\BibitemOpen
	\bibfield  {author} {\bibinfo {author} {\bibfnamefont {S.}~\bibnamefont
			{Massar}}\ and\ \bibinfo {author} {\bibfnamefont {S.}~\bibnamefont
			{Popescu}},\ }\href {\doibase 10.1103/PhysRevLett.74.1259} {\bibfield
		{journal} {\bibinfo  {journal} {Phys. Rev. Lett.}\ }\textbf {\bibinfo
			{volume} {74}},\ \bibinfo {pages} {1259} (\bibinfo {year}
		{1995})}\BibitemShut {NoStop}%
	\bibitem [{\citenamefont {Hou}\ \emph {et~al.}(2018)\citenamefont {Hou},
		\citenamefont {Tang}, \citenamefont {Shang}, \citenamefont {Zhu},
		\citenamefont {Li}, \citenamefont {Yuan}, \citenamefont {Wu}, \citenamefont
		{Xiang}, \citenamefont {Li},\ and\ \citenamefont
		{Guo}}]{hou2018deterministic}%
	\BibitemOpen
	\bibfield  {author} {\bibinfo {author} {\bibfnamefont {Z.}~\bibnamefont
			{Hou}}, \bibinfo {author} {\bibfnamefont {J.-F.}\ \bibnamefont {Tang}},
		\bibinfo {author} {\bibfnamefont {J.}~\bibnamefont {Shang}}, \bibinfo
		{author} {\bibfnamefont {H.}~\bibnamefont {Zhu}}, \bibinfo {author}
		{\bibfnamefont {J.}~\bibnamefont {Li}}, \bibinfo {author} {\bibfnamefont
			{Y.}~\bibnamefont {Yuan}}, \bibinfo {author} {\bibfnamefont {K.-D.}\
			\bibnamefont {Wu}}, \bibinfo {author} {\bibfnamefont {G.-Y.}\ \bibnamefont
			{Xiang}}, \bibinfo {author} {\bibfnamefont {C.-F.}\ \bibnamefont {Li}}, \
		and\ \bibinfo {author} {\bibfnamefont {G.-C.}\ \bibnamefont {Guo}},\ }\href
	{https://www.nature.com/articles/s41467-018-03849-x} {\bibfield  {journal}
		{\bibinfo  {journal} {Nature communications}\ }\textbf {\bibinfo {volume}
			{9}},\ \bibinfo {pages} {1414} (\bibinfo {year} {2018})}\BibitemShut
	{NoStop}%
	\bibitem [{\citenamefont {Cox}\ \emph {et~al.}(2016)\citenamefont {Cox},
		\citenamefont {Greve}, \citenamefont {Weiner},\ and\ \citenamefont
		{Thompson}}]{Cox2016}%
	\BibitemOpen
	\bibfield  {author} {\bibinfo {author} {\bibfnamefont {K.~C.}\ \bibnamefont
			{Cox}}, \bibinfo {author} {\bibfnamefont {G.~P.}\ \bibnamefont {Greve}},
		\bibinfo {author} {\bibfnamefont {J.~M.}\ \bibnamefont {Weiner}}, \ and\
		\bibinfo {author} {\bibfnamefont {J.~K.}\ \bibnamefont {Thompson}},\ }\href
	{\doibase 10.1103/PhysRevLett.116.093602} {\bibfield  {journal} {\bibinfo
			{journal} {Phys. Rev. Lett.}\ }\textbf {\bibinfo {volume} {116}},\ \bibinfo
		{pages} {093602} (\bibinfo {year} {2016})}\BibitemShut {NoStop}%
	\bibitem [{\citenamefont {Kammerlander}\ and\ \citenamefont
		{Anders}(2016)}]{Kammerlander2016quantum}%
	\BibitemOpen
	\bibfield  {author} {\bibinfo {author} {\bibfnamefont {P.}~\bibnamefont
			{Kammerlander}}\ and\ \bibinfo {author} {\bibfnamefont {J.}~\bibnamefont
			{Anders}},\ }\href {\doibase 10.1038/srep22174} {\bibfield  {journal}
		{\bibinfo  {journal} {Scientific Reports}\ }\textbf {\bibinfo {volume} {6}},\
		\bibinfo {pages} {22174} (\bibinfo {year} {2016})}\BibitemShut {NoStop}%
	\bibitem [{\citenamefont {Nielsen}\ and\ \citenamefont
		{Chuang}(2002)}]{nielsen2002quantum}%
	\BibitemOpen
	\bibfield  {author} {\bibinfo {author} {\bibfnamefont {M.~A.}\ \bibnamefont
			{Nielsen}}\ and\ \bibinfo {author} {\bibfnamefont {I.}~\bibnamefont
			{Chuang}},\ }\href@noop {} {\enquote {\bibinfo {title} {Quantum computation
				and quantum information},}\ } (\bibinfo {year} {2002})\BibitemShut {NoStop}%
	\bibitem [{\citenamefont {Takeuchi}(2001)}]{takeuchi2001beamlike}%
	\BibitemOpen
	\bibfield  {author} {\bibinfo {author} {\bibfnamefont {S.}~\bibnamefont
			{Takeuchi}},\ }\href@noop {} {\bibfield  {journal} {\bibinfo  {journal}
			{Optics Letters}\ }\textbf {\bibinfo {volume} {26}},\ \bibinfo {pages} {843}
		(\bibinfo {year} {2001})}\BibitemShut {NoStop}%
	\bibitem [{\citenamefont {Baumgratz}\ \emph {et~al.}(2014)\citenamefont
		{Baumgratz}, \citenamefont {Cramer},\ and\ \citenamefont
		{Plenio}}]{QuantifyingCoherence}%
	\BibitemOpen
	\bibfield  {author} {\bibinfo {author} {\bibfnamefont {T.}~\bibnamefont
			{Baumgratz}}, \bibinfo {author} {\bibfnamefont {M.}~\bibnamefont {Cramer}}, \
		and\ \bibinfo {author} {\bibfnamefont {M.~B.}\ \bibnamefont {Plenio}},\
	}\href {\doibase 10.1103/PhysRevLett.113.140401} {\bibfield  {journal}
		{\bibinfo  {journal} {Phys. Rev. Lett.}\ }\textbf {\bibinfo {volume} {113}},\
		\bibinfo {pages} {140401} (\bibinfo {year} {2014})}\BibitemShut {NoStop}%
	\bibitem [{\citenamefont {Bu}\ \emph {et~al.}(2017)\citenamefont {Bu},
		\citenamefont {Kumar}, \citenamefont {Zhang},\ and\ \citenamefont
		{Wu}}]{coheringpowerPLA}%
	\BibitemOpen
	\bibfield  {author} {\bibinfo {author} {\bibfnamefont {K.}~\bibnamefont
			{Bu}}, \bibinfo {author} {\bibfnamefont {A.}~\bibnamefont {Kumar}}, \bibinfo
		{author} {\bibfnamefont {L.}~\bibnamefont {Zhang}}, \ and\ \bibinfo {author}
		{\bibfnamefont {J.}~\bibnamefont {Wu}},\ }\href
	{https://www.sciencedirect.com/science/article/pii/S0375960117302621?via%3Dihub}
		{\bibfield  {journal} {\bibinfo  {journal} {Physics Letters A}\ }\textbf
			{\bibinfo {volume} {381}},\ \bibinfo {pages} {1670} (\bibinfo {year}
			{2017})}\BibitemShut {NoStop}%
		\bibitem [{\citenamefont {Alhambra}\ \emph {et~al.}(2016)\citenamefont
			{Alhambra}, \citenamefont {Masanes}, \citenamefont {Oppenheim},\ and\
			\citenamefont {Perry}}]{PhysRevX.6.041017}%
		\BibitemOpen
		\bibfield  {author} {\bibinfo {author} {\bibfnamefont {A.~M.}\ \bibnamefont
				{Alhambra}}, \bibinfo {author} {\bibfnamefont {L.}~\bibnamefont {Masanes}},
			\bibinfo {author} {\bibfnamefont {J.}~\bibnamefont {Oppenheim}}, \ and\
			\bibinfo {author} {\bibfnamefont {C.}~\bibnamefont {Perry}},\ }\href
		{\doibase 10.1103/PhysRevX.6.041017} {\bibfield  {journal} {\bibinfo
				{journal} {Phys. Rev. X}\ }\textbf {\bibinfo {volume} {6}},\ \bibinfo {pages}
			{041017} (\bibinfo {year} {2016})}\BibitemShut {NoStop}%
		\bibitem [{\citenamefont {\AA{}berg}(2018)}]{PhysRevX.8.011019}%
		\BibitemOpen
		\bibfield  {author} {\bibinfo {author} {\bibfnamefont {J.}~\bibnamefont
				{\AA{}berg}},\ }\href {\doibase 10.1103/PhysRevX.8.011019} {\bibfield
			{journal} {\bibinfo  {journal} {Phys. Rev. X}\ }\textbf {\bibinfo {volume}
				{8}},\ \bibinfo {pages} {011019} (\bibinfo {year} {2018})}\BibitemShut
		{NoStop}%
		\bibitem [{\citenamefont {Miller}\ and\ \citenamefont
			{Anders}(2018)}]{miller2018leggett}%
		\BibitemOpen
		\bibfield  {author} {\bibinfo {author} {\bibfnamefont {H.~J.}\ \bibnamefont
				{Miller}}\ and\ \bibinfo {author} {\bibfnamefont {J.}~\bibnamefont
				{Anders}},\ }\href@noop {} {\bibfield  {journal} {\bibinfo  {journal}
				{Entropy}\ }\textbf {\bibinfo {volume} {20}},\ \bibinfo {pages} {200}
			(\bibinfo {year} {2018})}\BibitemShut {NoStop}%
		\bibitem [{\citenamefont {Huber}\ \emph {et~al.}(2008)\citenamefont {Huber},
			\citenamefont {Schmidt-Kaler}, \citenamefont {Deffner},\ and\ \citenamefont
			{Lutz}}]{Huber2008employing}%
		\BibitemOpen
		\bibfield  {author} {\bibinfo {author} {\bibfnamefont {G.}~\bibnamefont
				{Huber}}, \bibinfo {author} {\bibfnamefont {F.}~\bibnamefont
				{Schmidt-Kaler}}, \bibinfo {author} {\bibfnamefont {S.}~\bibnamefont
				{Deffner}}, \ and\ \bibinfo {author} {\bibfnamefont {E.}~\bibnamefont
				{Lutz}},\ }\href {\doibase 10.1103/PhysRevLett.101.070403} {\bibfield
			{journal} {\bibinfo  {journal} {Phys. Rev. Lett.}\ }\textbf {\bibinfo
				{volume} {101}},\ \bibinfo {pages} {070403} (\bibinfo {year}
			{2008})}\BibitemShut {NoStop}%
		\bibitem [{\citenamefont {Batalh\~ao}\ \emph {et~al.}(2014)\citenamefont
			{Batalh\~ao}, \citenamefont {Souza}, \citenamefont {Mazzola}, \citenamefont
			{Auccaise}, \citenamefont {Sarthour}, \citenamefont {Oliveira}, \citenamefont
			{Goold}, \citenamefont {De~Chiara}, \citenamefont {Paternostro},\ and\
			\citenamefont {Serra}}]{Tiago2014experimental}%
		\BibitemOpen
		\bibfield  {author} {\bibinfo {author} {\bibfnamefont {T.~B.}\ \bibnamefont
				{Batalh\~ao}}, \bibinfo {author} {\bibfnamefont {A.~M.}\ \bibnamefont
				{Souza}}, \bibinfo {author} {\bibfnamefont {L.}~\bibnamefont {Mazzola}},
			\bibinfo {author} {\bibfnamefont {R.}~\bibnamefont {Auccaise}}, \bibinfo
			{author} {\bibfnamefont {R.~S.}\ \bibnamefont {Sarthour}}, \bibinfo {author}
			{\bibfnamefont {I.~S.}\ \bibnamefont {Oliveira}}, \bibinfo {author}
			{\bibfnamefont {J.}~\bibnamefont {Goold}}, \bibinfo {author} {\bibfnamefont
				{G.}~\bibnamefont {De~Chiara}}, \bibinfo {author} {\bibfnamefont
				{M.}~\bibnamefont {Paternostro}}, \ and\ \bibinfo {author} {\bibfnamefont
				{R.~M.}\ \bibnamefont {Serra}},\ }\href {\doibase
			10.1103/PhysRevLett.113.140601} {\bibfield  {journal} {\bibinfo  {journal}
				{Phys. Rev. Lett.}\ }\textbf {\bibinfo {volume} {113}},\ \bibinfo {pages}
			{140601} (\bibinfo {year} {2014})}\BibitemShut {NoStop}%
		\bibitem [{\citenamefont {An}\ \emph {et~al.}(2014)\citenamefont {An},
			\citenamefont {Zhang}, \citenamefont {Um}, \citenamefont {Lv}, \citenamefont
			{Lu}, \citenamefont {Zhang}, \citenamefont {Yin}, \citenamefont {Quan},\ and\
			\citenamefont {Kim}}]{Shuoming2014experimental}%
		\BibitemOpen
		\bibfield  {author} {\bibinfo {author} {\bibfnamefont {S.}~\bibnamefont
				{An}}, \bibinfo {author} {\bibfnamefont {J.-N.}\ \bibnamefont {Zhang}},
			\bibinfo {author} {\bibfnamefont {M.}~\bibnamefont {Um}}, \bibinfo {author}
			{\bibfnamefont {D.}~\bibnamefont {Lv}}, \bibinfo {author} {\bibfnamefont
				{Y.}~\bibnamefont {Lu}}, \bibinfo {author} {\bibfnamefont {J.}~\bibnamefont
				{Zhang}}, \bibinfo {author} {\bibfnamefont {Z.-Q.}\ \bibnamefont {Yin}},
			\bibinfo {author} {\bibfnamefont {H.~T.}\ \bibnamefont {Quan}}, \ and\
			\bibinfo {author} {\bibfnamefont {K.}~\bibnamefont {Kim}},\ }\href {\doibase
			10.1038/nphys3197} {\bibfield  {journal} {\bibinfo  {journal} {Nature
					Physics}\ }\textbf {\bibinfo {volume} {11}},\ \bibinfo {pages} {193}
			(\bibinfo {year} {2014})}\BibitemShut {NoStop}%
		\bibitem [{\citenamefont {Cerisola}\ \emph {et~al.}(2017)\citenamefont
			{Cerisola}, \citenamefont {Margalit}, \citenamefont {Machluf}, \citenamefont
			{Roncaglia}, \citenamefont {Paz},\ and\ \citenamefont
			{Folman}}]{Cerisola2017quantum}%
		\BibitemOpen
		\bibfield  {author} {\bibinfo {author} {\bibfnamefont {F.}~\bibnamefont
				{Cerisola}}, \bibinfo {author} {\bibfnamefont {Y.}~\bibnamefont {Margalit}},
			\bibinfo {author} {\bibfnamefont {S.}~\bibnamefont {Machluf}}, \bibinfo
			{author} {\bibfnamefont {A.~J.}\ \bibnamefont {Roncaglia}}, \bibinfo {author}
			{\bibfnamefont {J.~P.}\ \bibnamefont {Paz}}, \ and\ \bibinfo {author}
			{\bibfnamefont {R.}~\bibnamefont {Folman}},\ }\href {\doibase
			10.1038/s41467-017-01308-7} {\bibfield  {journal} {\bibinfo  {journal}
				{Nature Communications}\ }\textbf {\bibinfo {volume} {8}},\ \bibinfo {pages}
			{1241} (\bibinfo {year} {2017})}\BibitemShut {NoStop}%
	\end{thebibliography}
\end{document}